\RequirePackage{graphicx}
\documentclass[prd, superscriptaddress, nofootinbib, floatfix]{revtex4-2}   %\documentclass[prd, twocolumn, superscriptaddress, nofootinbib, floatfix]{revtex4-2}  

\usepackage{xcolor}
\usepackage[subfigure]{graphfig}
\usepackage[normalem]{ulem} %to strike the words
\usepackage{wrapfig}
\usepackage{amsmath}
\usepackage{amsfonts}
\usepackage{amssymb}
\usepackage{booktabs}
\usepackage{multirow}
\usepackage{slashed}
\usepackage{physics}
\usepackage{tabularx}
\usepackage{soul}
\usepackage{ulem}
\usepackage{hhline}
\usepackage{float}
\usepackage[colorlinks=true, linkcolor=blue, citecolor=blue, urlcolor=blue]{hyperref}
\usepackage{longtable}

\usepackage{nccmath}
\usepackage{esvect}

\newcommand\befs{\begin{figure*}}
\newcommand\eefs[1]{\label{fig:#1}\end{figure*}}
\newcommand\bef{\begin{figure}}
\newcommand\eef[1]{\label{fig:#1}\end{figure}}
\newcommand\beq{\begin{equation}}
\newcommand\eeq[1]{\label{#1}\end{equation}}
\newcommand\beqa{\begin{eqnarray}}
\newcommand\eeqa[1]{\label{#1}\end{eqnarray}}
\newcommand\bet{\begin{table}}
\newcommand\eet[1]{\label{tb:#1}\end{table}}
\newcommand\bets{\begin{table*}}
\newcommand\eets[1]{\label{tb:#1}\end{table*}}

\def\be{\begin{equation}}
\def\ee{\end{equation}}
\newcommand{\bea}{\begin{eqnarray}}
\newcommand{\eea}{\end{eqnarray}}

\newcommand{\nn}{\nonumber}

\definecolor{amethyst}{rgb}{0.6, 0.4, 0.8}

\newcommand{\ms}{\overline{\rm{MS}}}

\begin{document}
\preprint{JLAB-THY-21-3469}
\widetext

\title{\Large Unpolarized gluon distribution in the nucleon from lattice quantum chromodynamics}
\newcommand*{\WM}{Department of Physics, William and Mary, Williamsburg, Virginia, USA.}\affiliation{\WM}       %1
\newcommand*{\Jlab}{Thomas Jefferson National Accelerator Facility, Newport News, Virginia, USA.}\affiliation{\Jlab}    %2
\newcommand*{\CU}{Department of Physics, Columbia University, New York City, New York, USA.}\affiliation{\CU}   %3
\newcommand*{\ORNL}{Oak Ridge National Laboratory, Oak Ridge, Tennessee, USA.}\affiliation{\ORNL} 
\newcommand*{\ODU}{Department of Physics, Old Dominion University, Norfolk, Virginia, USA.}\affiliation{\ODU}    %4
\newcommand*{\CNRS}{Aix Marseille Univ, Universit\'e de Toulon, CNRS, CPT, Marseille, France.}\affiliation{\CNRS}    %4

%%%%%%%%%%%%%%%%%%%%%%%%%%%
% AUTH BLOCK
%%%%%%%%%%%%%%%%%%%%%%%%%%%
\author{Tanjib Khan}\affiliation{\WM}
\author{Raza Sabbir Sufian}\affiliation{\WM}\affiliation{\Jlab}
\author{Joseph Karpie}\affiliation{\CU}
\author{Christopher J. Monahan}\affiliation{\WM}\affiliation{\Jlab}
\author{Colin Egerer}\affiliation{\WM}\affiliation{\Jlab}
\author{B\'alint Jo\'o}\affiliation{\ORNL}
\author{Wayne Morris}\affiliation{\ODU}\affiliation{\Jlab}
\author{Kostas Orginos}\affiliation{\WM}\affiliation{\Jlab}
\author{Anatoly  Radyushkin}\affiliation{\ODU}\affiliation{\Jlab}
\author{David G. Richards}\affiliation{\Jlab} 
\author{Eloy Romero}\affiliation{\Jlab}
\author{Savvas Zafeiropoulos}\affiliation{\CNRS}
\collaboration{On behalf of the \textit{HadStruc Collaboration}}

\begin{abstract}
In this study, we present a determination of the unpolarized gluon Ioffe-time distribution in the nucleon from a first principles lattice quantum chromodynamics calculation. We carry out the  lattice calculation on a $32^3\times 64$ ensemble with a pion mass of $358$ MeV and lattice spacing of $0.094$ fm. We construct the nucleon interpolating fields using the distillation technique, flow the gauge fields using the gradient flow, and solve the summed generalized eigenvalue problem to determine the gluonic matrix elements. Combining these techniques allows us to provide a statistically well-controlled Ioffe-time distribution and unpolarized gluon PDF.  We obtain the flow time independent reduced Ioffe-time pseudo-distribution, and calculate the light-cone Ioffe-time distribution and unpolarized gluon distribution function in the $\ms$ scheme at $\mu = 2$ GeV,  neglecting the mixing of the gluon operator with the quark singlet sector. Finally, we compare our results to phenomenological determinations. 
\end{abstract}
%\date{\today}
\maketitle

%%%%%
\section{Introduction}\label{sec:intro}
%%%%%
Gluons, which carry color charge and serve as the mediator bosons of the strong interaction, play a key role in the nucleon's mass and spin. Confinement in quantum chromodynamics (QCD) ensures that no free quarks or gluons have been observed, so analyses of hadrons involving high energy scattering rely on QCD factorization~\cite{Collins:1989gx}. Factorization separates the perturbatively-calculable hard-scattering quark and gluon dynamics from the non-perturbative collinear dynamics, described by parton distribution functions (PDFs) of the relevant hadrons.

There are long-standing efforts to conduct global analyses~\cite{Bailey:2020ooq, Hou:2019efy, Ball:2017nwa, Accardi:2016qay, Dulat:2015mca} of data from available deep inelastic scattering (DIS) and related hard scattering processes to explore the nature of the PDFs. 
It is essential to have a clear and precise understanding of the gluon PDF in order to calculate the cross-section for Higgs boson production~\cite{CMS:2012nga} and jet production~\cite{Asquith:2018igt} at the Large Hadron Collider (LHC), and $J/\psi$ photo production~\cite{JLab:J_psi} at Jefferson Lab. Future colliders, such as the Electron Ion Collider (EIC)~\cite{Accardi:2012qut, Aguilar:2019teb, AbdulKhalek:2021gbh}, which is to be built at Brookhaven National Lab, and the Electron Ion Collider in China (EicC)~\cite{Anderle:2021wcy}, are expected to make significant impact on the precision of the gluon PDFs.  
While the precision of the extracted gluon distribution $x\,g(x)$ has been improved over the last decade, several issues remain unresolved; for example, the suppression in the momentum fraction region $ 0.1 < x < 0.4$ when ATLAS and CMS jet data are included~\cite{Hou:2019efy} and how to obtain a more precise determination of $g(x)$ are subjects of ongoing efforts. 

The determination of PDFs from lattice QCD is of particular theoretical interest to directly explore the non-perturbative sector of QCD from the first principles. To achieve this goal, there have been several proposals for the extraction of the $x$-dependent hadron structure from lattice QCD calculations, such as the path-integral formulation of the deep-inelastic scattering hadronic tensor~\cite{Liu:1993cv}, the operator product expansion~\cite{Detmold:2005gg}, quasi-PDFs~\cite{Ji:2013dva, Ji:2014gla},  pseudo-PDFs~\cite{Radyushkin:2017cyf}, and lattice cross-sections~\cite{Ma:2014jla,Ma:2017pxb}.  Lattice QCD is formulated in Euclidean space, so the bilocal light-cone correlators that are necessary to extract the PDFs cannot be evaluated directly, because they require operators containing fields at light-like separations, $z^2=0$, which cannot exist in Euclidean space. The quasi-PDF framework~\cite{Ji:2013dva} circumvents this drawback by calculating matrix elements associated with equal time and purely space-like field separations with hadron states at non-zero momentum, $p_z$. The corresponding quasi-PDFs can be matched to the light-cone PDFs when the hadron momentum is large, by applying the Large Momentum Effective Theory (LaMET)~\cite{Ji:2014gla}. These calculation techniques have been explored extensively in numerical lattice calculations. (For recent reviews see~\cite{Constantinou:2020hdm, Cichy:2018mum} and the references therein.)

There have been significant achievements in lattice QCD calculations of $x$-dependent hadron structure: the nucleon valence quark distribution using  pseudo-PDFs~\cite{Karpie:2021pap}, the calculation of the pion valence distribution using the lattice cross section, quasi-PDF and pseudo-PDF frameworks~\cite{Sufian:2020vzb, Sufian:2019bol, Zhang:2020rsx, Izubuchi:2019lyk,  Gao:2020ito}, the kaon PDF  calculation using the quasi-PDF formalism~\cite{Lin:2020ssv}, nucleon unpolarized and helicity distributions within quasi-PDF formalism~\cite{Alexandrou:2020qtt, Alexandrou:2020uyt, Fan:2020nzz}, the unpolarized and helicity GPD calculation of the proton~\cite{Alexandrou:2020zbe}, and a quasi-TMD calculation in the pion~\cite{LatticeParton:2020uhz}.  However, there are fewer lattice calculations of gluon distribution functions than that of quark distributions. Lattice calculations include the gluon momentum fraction~\cite{Alexandrou:2020sml, Yang:2018nqn}, the gluon contribution to the nucleon spin~\cite{Alexandrou:2017oeh}, gluon gravitational form factors of the nucleon and the pion~\cite{Shanahan:2018pib}. Recently, there have been attempts to calculate gluon PDFs in the nucleon~\cite{Fan:2018dxu,Fan:2020cpa} and in the pion~\cite{Fan:2021bcr}.

In this work, we apply the pseudo-PDF approach~\cite{Radyushkin:2017cyf} to extract the gluon PDF in the nucleon. We calculate the Ioffe-time pseudo-distribution function (pseudo-ITD), $\mathfrak{M} (\nu, z^2)$~\cite{Radyushkin:2017cyf, Radyushkin:2017lvu, Radyushkin:2018cvn}, where the Ioffe-time~\cite{Ioffe:1969kf} is a dimensionless quantity that describes the length of time that the DIS probe interacts with the nucleon, in units of the inverse hadron mass. The related pseudo-PDF, $\mathcal{P} (x, z^2)$ can be determined from the Fourier transform of the pseudo-ITD. The pseudo-PDF and the pseudo-ITD are the Lorentz invariant generalizations of the PDF and of the Ioffe-time distribution function (ITD)~\cite{Braun:1994jq} to non-zero separations, $z^2>0$, respectively. In renormalizable theories, the pseudo-PDF has a logarithmic divergence at small $z$-separations that corresponds to the DGLAP evolution of the PDF. The pseudo-PDF and the pseudo-ITD can be factorized into the PDF and perturbatively calculable kernels, similar to the factorization framework for experimental cross-sections. There have been a number of lattice calculations implementing the pseudo-PDF method~\cite{Orginos:2017kos, Karpie:2018zaz,Joo:2019jct,Joo:2019bzr, Joo:2020spy,Bhat:2020ktg}. Our calculation applies the reduced pseudo-ITD approach, in which the multiplicative UV renormalization factors are canceled by constructing a ratio of the relevant matrix elements~\cite{Joo:2019jct}. This ratio, the reduced pseudo-ITD, removes the Wilson-line related divergences, as well as various other systematic errors.  We determine the gluon PDF from the reduced pseudo-ITD through the short distance factorization (SDF).

The unpolarized gluon PDF must be extracted from our lattice results by inverting the convolution that relates the PDF to the lattice matrix elements. We have access to a limited number of discrete and noisy values of the matrix element on the lattice, so this inversion problem is ill-posed. A number of techniques have been proposed to overcome this inverse problem~\cite{Karpie:2019eiq}, such as discrete Fourier transform, the Backus-Gilbert method~\cite{Karpie:2019eiq, Bhat:2020ktg}, the Bayes-Gauss-Fourier transform~\cite{Alexandrou:2020qtt}, adapting phenomenologically-motivated functional forms~\cite{Sufian:2020vzb}, and finally the application of neural networks~\cite{Cichy:2019ebf,DelDebbio:2020rgv}, which provide more flexible parameterizations of the PDFs. Here, we parameterize the reduced pseudo-ITD using Jacobi polynomials~\cite{Karpie:2021pap, Egerer:2021ymv}. We vary the parameterization of the lattice matrix elements to incorporate different correction terms and to compare multiple functional forms for the gluon PDF to study the parameterization dependence.

The rest of this paper is organized as follows. In Sec.~\ref{sec:theory}, we first identify the matrix elements needed to calculate the unpolarized gluon parton distribution, construct the reduced pseudo-ITD from the matrix elements and lay out the position-space matching that relates the reduced pseudo-ITD to the light-cone ITD. In Sec.~\ref{comp_frame}, we describe the construction of the gluonic currents associated with the matrix elements and the nucleon two-point correlators. Sec.~\ref{sec:latt_details} contains the details of our lattice setup. In Sec.~\ref{var_analysis}, we demonstrate the consistency of the nucleon two-point correlators by extracting the energy spectra. Sec.~\ref{sec:mtx_calc} describes the methodology we implement to calculate the reduced pseudo-ITD from the three-point correlators. In Sec.~\ref{sec:pdf_calc}, we extract the gluon PDF from the reduced pseudo-ITD and compare our results with the phenomenological distributions. Sec.~\ref{conclusion} contains our concluding remarks.

%%%%%
%%%%%
%%%%%

\section{Theoretical Background of Gluon pseudo-Distributions }\label{sec:theory}
\subsection{Matrix Elements}\label{sec:mtx_elem}

To access the unpolarized gluon PDF, we calculate the matrix elements of a spin-averaged nucleon for operators composed of two gluon fields connected by a Wilson line, which have the general form
\bea
    M_{\mu \alpha; \lambda \beta} (z, p) \equiv \langle p| \, G_{\mu \alpha} (z) \, W[z, 0] \, G_{\lambda \beta} (0) \, |p\rangle \, .
\eea
Here, $z_\mu$ is the separation between the gluon-fields, $p_\mu$ is the 4-momentum of the nucleon, $W[z, 0]$ is the standard straight-line Wilson line in the adjoint representation,
\bea
    W[x, y] = {\cal P}\mathrm{exp} \Big\{ ig_s \int_0^1 d\eta \, (x-y)^\mu \Tilde{A}_\mu \big(\eta x + (1-\eta)y \big) \Big\} \, ,
\eea
for the gauge field $A_\mu$, where ${\cal P}$ indicates that the integral is path-ordered. The matrix elements can be decomposed into invariant amplitudes, $\mathcal{M}_{pp}$, $\mathcal{M}_{zz}$, $\mathcal{M}_{zp}$, $\mathcal{M}_{pz}$, $\mathcal{M}_{ppzz}$ and $\mathcal{M}_{gg}$ using the four-vectors, $p_\mu$ and $z_\mu$, and the metric tensor $g_{\mu \nu}$~\cite{Balitsky:2019krf}. These amplitudes are functions of the invariant interval $z^2$ and the Ioffe-time $p \cdot z \equiv - \nu$~\cite{Ioffe:1969kf}.

The light-cone gluon distribution is obtained from
\bea
    g^{\alpha \beta} \, M_{+\alpha; \beta+} (z_-, p) = - 2 p_+^2 \, \mathcal{M}_{pp} (\nu, 0) \, ,\label{eq:lc_mpp}
\eea
where $z$ is taken in the light-cone ``minus'' direction, $z = z_-$, and $p^+$ is the momentum in the light-cone ``plus'' direction. The PDF is determined by the $\mathcal{M}_{pp}$ amplitude,
\bea
    -\mathcal{M}_{pp} (\nu, 0) = \frac{1}{2} \int_{-1}^1 \mathrm{d} x \, e^{-ix \nu} \, x\,g(x) \, .
\eea

The density of the momentum carried by the gluons, $\mathcal{G}(x) = x\,g(x)$ is the natural quantity in this definition of the gluon PDF, rather than $g(x)$. The field-strength tensor $G_{\mu \alpha}$ is antisymmetric with respect to its indices and $g_{--} = 0$, so the left hand side of Eq.~\eqref{eq:lc_mpp} reduces to a summation over the transverse indices $i,j = x, y$; perpendicular to the direction of separation between the two gluon fields. The matrix element $M_{ti;it}$ decomposes into the invariant amplitudes~\cite{Balitsky:2019krf}
\bea
    M_{ti;it} = 2\, p_0^2 \mathcal{M}_{pp} + 2\mathcal{M}_{gg} \, ,
\eea
where $\mathcal{M}_{gg}$ is a contamination term. The matrix element
\bea
    M_{ji;ij} = \langle p| \, G_{ji} (z) \, W[z, 0] \, G_{ij} (0) \, |p\rangle = -2 \mathcal{M}_{gg} \, ,
\eea
cancels the contamination term from $M_{ti;it}$~\cite{Balitsky:2019krf}. Thus, the proper combination of the matrix elements to extract the twist-2 invariant amplitude, $\mathcal{M}_{pp}$ is
\bea 
    M_{ti;it} + M_{ji;ij} = 2 p_0^2 \, \mathcal{M}_{pp} \, . \label{eq:pseudo_mpp}
\eea

For spatially-separated fields, the gauge link operator has extra ultraviolet divergences not present for light-like separated fields. The combination of matrix elements $M_{ti;it}$ is multiplicatively renormalizable~\cite{Zhang:2018diq}.  And, because of the antisymmetry of the gluon fields, the combination $M_{ji;ij}$ can be written as
\bea
    M_{ji;ij} = 2 \, \langle p| \, G_{yx} (z) \, W[z, 0] \, G_{xy} (0) \, |p\rangle \, ,
\eea
which contains only one set of indices $\{\mu \alpha; \lambda \beta\}$, making explicit the fact that 
this matrix element is multiplicatively renormalizable too~\cite{Li:2018tpe}. Furthermore, both $M_{ti;it}$ and $M_{ji;ij}$ have the same one-loop UV anomalous dimension~\cite{Balitsky:2019krf}, making the whole combination in Eq.~\eqref{eq:pseudo_mpp} multiplicatively renormalizable at the one-loop level, at least.

%%%%%%%%%%%%%%%%%%%%%%%
%%%%%%%%%%%%%%%%%%%%%%%
\subsection{Reduced Matrix Elements}\label{sec:reduced}
Similar to space-like separations, the extended gluon operator has additional link-related ultraviolet (UV) divergences which are multiplicatively renormalizable~\cite{Izubuchi:2018srq,Ji:2017oey,Green:2017xeu}. These UV divergences can be cancelled by taking appropriate ratios. We combine the matrix elements from Eq.~\eqref{eq:pseudo_mpp} which we denote by $\mathcal{M}(\nu,z^2)$ for the rest of the paper, and take the ratio~\cite{Orginos:2017kos} of the combination to its rest-frame value, keeping the separation same. This ratio cancels out the $\nu$-independent UV factor $Z(z^2/a^2)$, making the ratio UV-finite. The kinematic factors remaining in the ratio can be removed by taking the ratio of the non-zero separation to the zero separation matrix elements, at fixed Ioffe-time, in both the numerator and denominator~\cite{Joo:2019jct}. 

The resulting reduced matrix element, the reduced pseudo-ITD, can be written as:
\bea \label{eq:doubleratio}
&& \mathfrak{M} (\nu, z^2) = \Bigg( \frac{\mathcal{M} (\nu, z^2 )}{\mathcal{M} (\nu, 0)|_{z=0}} \Bigg) / \Bigg( \frac{\mathcal{M} (0, z^2)|_{p=0}}{\mathcal{M} (0, 0)|_{p=0, z=0}} \Bigg) \, .
\eea

Taking the ratio,  we also eliminate  
$z^2$-dependent, but  
$\nu$-independent, non-perturbative  factors  that 
$\mathcal{M} (\nu, z^2)$ may contain.
The residual polynomial ``higher twist'' dependence on $z^2$, if visible,
should be  explicitly  fitted in order to separate it  from
the twist-2 contribution.

%%%%%%%%%%%%%%%%%%%%%%%
%%%%%%%%%%%%%%%%%%%%%%%

\subsection{Position-space Matching}\label{sec:matching}

The reduced pseudo-ITD has a logarithmic $z^2$ dependence. We relate the reduced pseudo-ITD, $\mathfrak{M} (\nu, z^2)$, to the gluon and singlet quark light-cone ITDs,  $\mathcal{I}_g (\nu, \mu^2)$ and $\mathcal{I}_S (\nu, \mu^2)$ in the $\ms$ scheme through the short distance factorization relationship with $z^2$ as the hard scale. Here, $\mathcal{I}_g (\nu, \mu^2)$ is related to the gluon PDF, $g(x, \mu^2)$, by
\be
    \mathcal{I}_g (\nu, \mu^2) = \frac{1}{2} \int_{-1}^{1} dx \, e^{ix \nu} \, x \, g(x, \mu^2) \, .
\ee

The product $x\,g(x, \mu^2)$ is an even function of $x$, so the real part of $\mathcal{I}_g (\nu, \mu^2)$ is given by the cosine transform of $x \, g(x, \mu^2)$, while its imaginary part vanishes. 
Neglecting the higher twist terms of $\mathcal{M}_{zz}$, $\mathcal{M}_{zp}$, $\mathcal{M}_{pz}$, $\mathcal{M}_{ppzz}$, and keeping just the $\mathcal{M}_{pp}$ term, the one-loop matching relation 
is~\cite{Balitsky:2019krf,Balitsky:2021bds},
\bea
\mathfrak{M} (\nu, z^2) &=& \, \frac{\mathcal{I}_g (\nu, \mu^2)}{\mathcal{I}_g (0, \mu^2)}
- \, \frac{\alpha_s N_c}{2 \pi} \int_0^1 du \; \frac{\mathcal{I}_g (u \nu, \mu^2)}{\mathcal{I}_g (0, \mu^2)} \;  \Bigg\{ \mathrm{ln}\bigg( \frac{z^2 \mu^2 e^{2 \gamma_E}}{4} \bigg) \; B_{gg} (u) + \; 4 \bigg[ \frac{u + \ln(\bar{u})}{\bar{u}} \bigg]_+ \nn \\
&&  + \; \frac{2}{3} \Big[ 1 - u^3 \Big]_+ \Bigg\} 
 \hspace{0pt} -\, \frac{\alpha_s C_F}{2 \pi} \, \mathrm{ln}\bigg( \frac{z^2 \mu^2 e^{2 \gamma_E}}{4} \bigg) \int_0^1 dw \;  \frac{\mathcal{I}_S (w \nu, \mu^2)}{\mathcal{I}_g (0, \mu^2)} \; \mathfrak{B}_{gq}(w) \, . 
\eea
The singlet quark Ioffe-time distribution $\mathcal{I}_S (\nu, \mu^2)$ is related to the singlet quark distribution, summed over quark flavors. The Altarelli-Parisi  kernel, $B_{gg} (u)$, is given by
\be
    B_{gg} (u) = 2 \Bigg[\frac{(1 - u \bar{u})^2}{1 - u}\Bigg]_+  \, ,
\ee
and the quark-gluon mixing kernel is given by
\be
    \mathfrak{B}_{gq}(w) = \Big[ 1 + (1-w)^2 \Big]_+ \, ,
\ee
where the plus-prescription is
\be
    \int_0^1 du \, \Big[ f(u) \Big]_+ \; g(u) = \int_0^1 du \, f(u) \, \Big[ g(u) - g(1) \Big]
\ee
and $\bar{u} \equiv (1 - u)$.
Here, $\gamma_E$ is the Euler–Mascheroni constant and $C_F$ is the quadratic Casimir operator in the fundamental representation. Determining the singlet quark Ioffe-time distribution requires evaluation of the disconnected diagrams, which involves the computationally demanding calculation of the trace of the all-to-all quark propagator~\cite{Gambhir:2016jul}, but contribute only a little to the matching. We neglect quark-gluon mixing in this calculation and implement the matching relation
\bea
 \mathfrak{M} (\nu, z^2) &=& \, \frac{\mathcal{I}_g (\nu, \mu^2)}{\mathcal{I}_g (0, \mu^2)} 
 - \, \frac{\alpha_s N_c}{2 \pi} \int_0^1 du \; \frac{\mathcal{I}_g (u \nu, \mu^2)}{\mathcal{I}_g (0, \mu^2)} \;  \Bigg\{ \mathrm{ln}\bigg( \frac{z^2 \mu^2 e^{2 \gamma_E}}{4} \bigg) \; B_{gg} (u) \nn \\
&& + \; 4 \bigg[ \frac{u + \ln(\bar{u})}{\bar{u}} \bigg]_+ + \; \frac{2}{3} \Big[ 1 - u^3 \Big]_+ \Bigg\} \, . 
 \label{eq:matching}
\eea

%%%%%%%%%%%%%%%%%%%%%%%%%
%%%%%%%%%%%%%%%%%%%%%%%%%
\section{Computational Framework}\label{comp_frame}
%%%%%%%%%%%%%%%%%%%%%%%%%
%%%%%%%%%%%%%%%%%%%%%%%%%

\subsection{Gluonic Current Calculation}\label{sec:glu_curr}
%%%%%%%%%%%%%%%%%%%%%%%%%
%%%%%%%%%%%%%%%%%%%%%%%%%

The gluonic currents, inserted into the nucleon to calculate the matrix elements, are not connected to the nucleon state by any quark propagator, so the currents are largely decoupled from the nucleon part of the calculation itself. As a result, on the lattice, we can calculate the gluonic currents and the nucleon two-point correlators separately and combine them together to obtain the three-point correlators from which we extract the matrix elements. On the lattice, the gluonic current can be written with the Wilson line in the fundamental representation as
\bea
&& O(G_{\mu \alpha}, G_{\lambda \beta}, z) \;  \equiv \;  G_{\mu \alpha}(z) \; U(z,0) \; G_{\lambda \beta}(0) \; U(0,z). \nn \\
&&
\eea

The field-strength tensor can be expressed in terms of the $(1 \times 1)$ plaquette operator, $U_{\mu \nu}^{(1 \times 1)}$, as~\cite{BilsonThompson:2002jk} 
\bea
    \frac{-i}{2} \bigg[ U_{\mu \nu}^{(1 \times 1)} \; -  U_{\mu \nu}^{(1 \times 1)\dagger} \; - \; \frac{1}{3} \mathrm{Tr} \Big( U_{\mu \nu}^{(1 \times 1)} - U_{\mu \nu}^{(1 \times 1)\dagger} \Big) \bigg]
     = g_s a^2 \Big[ G_{\mu \nu} + \mathcal{O}(a^2) + \mathcal{O}(g_s^2 a^2) \Big],
\eea 
where $a$ is the lattice spacing and $\beta = 6/g_s^2$. One-third of the trace is subtracted here to enforce the traceless property of the Gell-Mann matrices. The $(1 \times 1)$ plaquette operator is defined as the product of the link variables forming a $(1 \times 1)$ loop on the lattice,
\be
U_{\mu \nu}^{(1 \times 1)} (x) = U_\mu (x) \, U_\nu (x+a\hat{\mu}) \, U_\mu^\dagger (x+a\hat{\nu}) \, U_\nu^\dagger (x).
\ee

\begin{figure}[t]
\center{\includegraphics[scale=0.8]{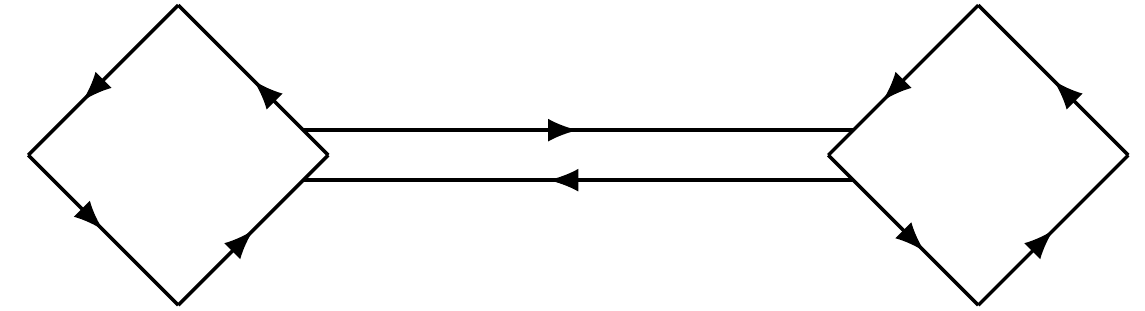}}
\caption{Visual representation of the gluonic current, $O(G_{\mu \alpha}, G_{\lambda \beta}, z)$. The rectangles on both the sides represent field-strength tensors and the lines connecting them represent the Wilson lines on the lattice. \label{fig:glu_current}}
\end{figure}

To reduce statistical fluctuations, we take the average of the four possible plaquette operators that can be constructed by changing the signs of $\mu$ and $\nu$. Finally, we combine the gluonic currents $O(G_{t i},G_{i t}, z)$ and $O(G_{j i},G_{i j}, z)$ to calculate $\mathcal{M}_{pp}$. Accounting for the sign change of the gluonic current with the ``temporal" index in Euclidean spacetime, the total gluonic current becomes
\bea
    O_{g}(z) =  G_{ji}(z) \, U(z,0) \, G_{ij}(0) \, U(0,z)
     - \; G_{ti}(z) \, U(z,0) \, G_{it}(0) \, U(0,z).
\eea

\subsection{Gradient Flow}\label{sec:gradient_flow}
%%%%%%%%%%%%%%%%%%%%%%%%%
%%%%%%%%%%%%%%%%%%%%%%%%%

In our calculation, we apply the gradient flow~\cite{Luscher:2010iy,Luscher:2011bx,Luscher:2013cpa} to reduce ultraviolet fluctuations and improve the signal-to-noise ratio for the gluon observables. To implement this technique, the flowed gauge field, $B_\mu (\tau, x)$, is defined by following the procedure in~\cite{Luscher:2010iy},
\bea
    &\;\;\; \dot{B}_\mu = D_{\nu} G_{\nu \mu}\;\;\;\;\;\;,\;\;\;\;\;\; D_\mu = \partial_\mu + [B_\mu, \,\cdot\;] \, , \nn \\
    & G_{\mu \nu} = \partial_\mu B_\nu - \partial_\nu B_\mu + [B_\mu, B_\nu] \, ,
\eea
where the flowed gauge field is subjected to the boundary condition $B_\mu(\tau=0, x) = A_\mu(x)$. Here $\tau$ is the flow time and we abbreviate differentiation with respect to $\tau$ by a dot. The flow equation of the gauge field is a diffusion equation and the evolution operator in the momentum space acts as an UV regulator for $\tau > 0$. As a result, the gradient flow exponentially suppresses the UV field-fluctuations, which corresponds to smearing out the original degrees of freedom in coordinate space. The operators constructed using flowed gauge fields with positive flow time enter into the relevant theories at length scales of $\sim\sqrt{8\tau}$. 

On the lattice, the gradient flow is implemented by defining the flowed link variable, $V_\mu (\tau, x)$ as~\cite{Luscher:2010iy}:
\bea
    \dot{V}_\mu (\tau, x) = - g_0^2 \{ \partial_{x, \mu} S(V_\mu(\tau, x)) \} V_\mu (\tau, x) \, ,
\eea
where $g_0$ is the bare coupling, $S(V_\mu(\tau, x))$ is the flowed action, $V_\mu (\tau = 0, x)$ has the boundary condition of being equal to the link variable, $U_\mu (x)$, and $\partial_{x, \mu}$ stands for the natural SU(3)-valued differential operator with respect to $V_\mu (\tau, x)$. The action, $S(V_\mu(\tau, x))$ is a monotonically decreasing function of $\tau$, and the gradient flow corresponds to a continuous stout-link smearing procedure~\cite{Morningstar:2003gk}.

We use unimproved Wilson flow and calculate the gluonic currents with flow times from $\tau = a^2$ to $\tau=3.8a^2$. We construct the double ratio of Eq.~\eqref{eq:doubleratio} using the flowed matrix elements, which further reduces UV fluctuations and suppresses the flow time dependence. The residual $\tau$-dependence is removed by fitting the flowed reduced matrix elements to an appropriate functional form which, in turn, gives us the reduced pseudo-ITD at zero flow time.

%%%%%%%%%%%%%%%%%%%%%%%%%
%%%%%%%%%%%%%%%%%%%%%%%%%
\subsection{Nucleon Two-point Correlator}\label{sec:2pt_corr}

We calculate the nucleon two-point correlators by applying interpolators at the source time-slice and the sink time-slice on the lattice. We apply distillation~\cite{Peardon:2009gh}, a low-rank approximation to the gauge-covariant Jacobi-smearing kernel, $J_{\sigma, n_\sigma} (t) = \Big( 1+\frac{\sigma \nabla^2 (t)}{n_\sigma} \Big)^{n_\sigma}$~\cite{Allton:1993wc}. The tunable parameters $\{ \sigma, n_\sigma \}$ ensure that, in the large iteration limit, the kernel approaches that of a spherically-symmetric Gaussian. The quark fields are smeared using the distillation smearing kernel
\be
    \square_{xy} (t) = \sum_{k = 1}^{N_D} \nu_x^{(k)} (t)\; \nu_y^{(k) \dagger} (t) \equiv V_D(t)\; V_D^\dagger (t),
\ee
where $V_D(t)$ is a $(N_c \times N_x \times N_y \times N_z) \times N_D$ matrix, where $N_c$ is the dimension of the color space, $N_x, N_y, N_z$ are the extents of the lattice in the three spatial directions, and $N_D$ is the dimension of the distillation space. The $k^\mathrm{th}$ column of $V_D(t)$, $\nu_x^{(k)} (t)$ is the $k^\mathrm{th}$ eigenvector of the second-order three-dimensional differential operator, $\nabla^2$, evaluated on the background of the spatial gauge fields of time-slice $t$, once the eigenvectors have been sorted by the ascending order of the eigenvalues. The two-point correlator for the nucleon can be written as
\bea
     \Big \langle \mathcal{O}_{N, i} (m) \bar{\mathcal{O}}_{N, j} (n) \Big \rangle  &=& \Phi^{(pqr)}_{i, \, \alpha \beta \gamma} (t_m) 
     \times \Big[ \mathfrak{P}^{p \bar{p}}_{\alpha \bar{\alpha}} (t_m, t_n) \; \mathfrak{P}^{q \bar{q}}_{\beta \bar{\beta}} (t_m, t_n) \; \mathfrak{P}^{r \bar{r}}_{\gamma \bar{\gamma}} (t_m, t_n) \nn \\
    && - \mathfrak{P}^{p \bar{p}}_{\alpha \bar{\alpha}} (t_m, t_n) \; \mathfrak{P}^{q \bar{r}}_{\beta \bar{\gamma}} (t_m, t_n) \; \mathfrak{P}^{r \bar{q}}_{\gamma \bar{\beta}} (t_m, t_n) \Big] 
     \times \Phi^{(\bar{p} \bar{q} \bar{r}) *}_{j, \, \bar{\alpha} \bar{\beta} \bar{\gamma}} (t_n),
\eea
where,
\bea
    \Phi^{(pqr)}_{i, \alpha \beta \gamma} (t) = \epsilon_{a b c} \; S_{i, \alpha \beta \gamma} \; \Big( \Gamma_{1i} \, \nu^{(p)} \Big)^a 
     \Big( \Gamma_{2i} \, \nu^{(q)} \Big)^b \; \Big( \Gamma_{3i} \, \nu^{(r)} \Big)^c (t),
\eea
and
\be
    \mathfrak{P}^{p \bar{p}}_{\alpha \bar{\alpha}} (t_m, t_n) = \nu^{(p)\dagger} (t_m) \; D^{-1}_{\alpha \bar{\alpha}} (t_m, t_n) \; \nu^{(\bar{p})} (t_n).
\ee
Here, $\Phi_i (t)$ and $\mathfrak{P}(t_m, t_n)$ are referred to as \textit{elementals} and \textit{perambulators}, respectively; $D$ is the lattice representation of the Dirac operator; $\alpha, \bar{\alpha}, \beta, \bar{\beta}, \gamma, \bar{\gamma}$ are the spin indices; $a$, $b$, $c$ are the color indices. The $\Phi_i (t)$ encodes the structure of the interpolating operator as well as has a well-defined momentum, while $\mathfrak{P}(t_m,t_n)$ encodes the propagation of the quarks, and does not have have any explicit momentum projection. Elementals can be decomposed into terms that act only within coordinate and color space, like $\Gamma$, and only within spin space, like $S_{\alpha \beta \gamma}$ .

We adopt distillation for two reasons. First, the computationally demanding parallel transporters of the theory, the perambulators, depend only on the gauge field, and not on the interpolators. Therefore, we can calculate the perambulators on an ensemble of gauge fields once, and then reuse them for an extended basis of interpolators, thus reducing the computational cost to a great extent. This extended basis of interpolators is the key to perform a successful summed generalized eigenvalue problem (sGEVP) analysis~\cite{Bulava:2011yz}, enabling us to attain a clear signal for the ground state nucleon.

Second, distillation admits a momentum projection both at the source interpolating operator, and at the sink interpolating operator, in contrast to the more usually adopted methods.  Thus for the gluonic three-point functions computed here, we are able to impose momentum projection at all three time-slices, ensuring the most complete possible sampling of the lattice. Moreover, the low-lying spectra of the nucleon can be faithfully captured with a relatively small number of distillation eigenvectors~\cite{Khan:2020ahz}, thus lowering the cost of the calculation further. The expectation is that $N_D$ should scale as the physical volume, and the cost of computing the corresponding correlation functions scales as $N_D^4$ for the case of the nucleon. In this calculation, we employed $N_D=64$ eigenvectors. The efficacy of distillation for the calculation of nucleon charges was demonstrated in ref.~\cite{Egerer:2018xgu}, and subsequently extended to the case of the nucleon in motion~\cite{Egerer:2020hnc}.  Recently, the unpolarized, isovector PDF of the nucleon has been computed using the same ensemble within the distillation framework~\cite{Egerer:2021ymv}.

\subsection{Interpolators}
%%%%%%%%%%%%%%%%%%%%%%%%%
%%%%%%%%%%%%%%%%%%%%%%%%%

The lattice regulator explicitly breaks the continuum  SO(3) rotational symmetry, so the associated symmetry group reduces to the double-cover octahedral group, $O^D_h$ for the nucleon at rest.  Although there are six irreducible representations (irreps.) available in $O^D_h$, we focus on $G_{1g}$, because the states with continuum spin $\frac{1}{2}$, such as the ground state nucleon, are subduced onto this irrep. Here, the subscript $g$ stands for positive parity. At non-zero spatial momenta, the $O^D_h$ group breaks into further little groups depending on the direction of the boost. We consider boosts only along the $z$-direction, so the associated little group is the order-16 dicyclic group or $\mathrm{Dic}_4$.
 
To calculate the low-lying spectra of the nucleon, we include interpolators with zero orbital angular momentum, which have the largest overlaps with the ground state of the nucleon. For the lowest excited-states, we include interpolators with gauge-covariant derivatives acting on the quark fields to capture the effect of the non-zero angular momenta between the quarks~\cite{Edwards:2011jj}. All these interpolators are ``non-relativistic", in the sense that they feature only the upper components of the Dirac spinors. We also include the interpolators that have derivatives of second order and form combinations corresponding to the commutation of two gauge-covariant derivatives acting on the same quark field. These interpolators, also referred to as hybrid interpolators~\cite{Dudek:2012ag}, vanish in the absence of a gauge-field and correspond to the chromomagnetic components of the gluonic field-strength tensor. We tabulate our choice of interpolators for the nucleon at rest as the first row in Table~\ref{tab:interpolator}, using the spectroscopic notation of: $X^{\;2S+1}L_\pi J^P$ where $X$ is the nucleon, $N$; $S$ is the Dirac spin; $L=S, \; P, \; D, \dots$ is the orbital angular momentum; $\pi = S, \; M \; \mathrm{or} \; A$ is the permutation symmetry of the derivatives; $J$ is the total angular momentum; and $P$ is the parity.  For the construction of the three-point correlators needed for the unpolarized distributions, we take the sum of the spin = +$\frac{1}{2}$ and spin = -$\frac{1}{2}$ nucleon two-point correlators.

For the case of the correlation functions at non-zero spatial momentum, parity is no longer a good quantum number and further operators are classified according to their helicity.  We therefore include operators corresponding both to higher spin, and to negative parity, in our basis within the little group $\mathrm{Dic}_4$. We choose the direction of momenta to be in the same direction of the polarization to ensure longitudinal polarization. We access the unpolarized gluon PDF by taking the sum of helicity = +$\frac{1}{2}$ and helicity = -$\frac{1}{2}$ nucleon two-point correlators.  The basis of interpolators is tabulated as the second row in Table~\ref{tab:interpolator}.

\begin{table}
  \renewcommand{\arraystretch}{1.5}
  \setlength{\tabcolsep}{10pt}
  \begin{tabular}{cc}
  \toprule
    Spatial momentum & Interpolators\\
    \midrule
    $\overrightarrow{p} = \overrightarrow{0}$ & $N^{\;2} S_S\, \frac{1}{2}^+, \;\; N^{\;2} S_M\, \frac{1}{2}^+, \;\; N^{\;4} D_M\, \frac{1}{2}^+$,\\
    & $N^{\;2} P_A\, \frac{1}{2}^+, \;\; N^{\;4} P_M^{\ast}\, \frac{1}{2}^+, \;\; N^{\;2} P_M^{\ast}\, \frac{1}{2}^+$\\
    \midrule
    $\overrightarrow{p} \neq \overrightarrow{0}$ & $N^{\;2} P_M\, \frac{1}{2}^-, \;\; N^{\;2} P_M\, \frac{3}{2}^-, \;\; N^{\;4} P_M\, \frac{1}{2}^-$,\\ 
    & $N^{\;4} P_M\, \frac{3}{2}^-, \;\; N^{\;4} P_M\, \frac{5}{2}^-, \;\; N^{\;2} S_S\, \frac{1}{2}^+$,\\
    & $N^{\;2} S_M\, \frac{1}{2}^+, \;\; N^{\;2} P_M^{\ast}\, \frac{1}{2}^+, \;\; N^{\;4} P_M^{\ast}\, \frac{1}{2}^+$\\
    \bottomrule
  \end{tabular}
  \caption{Nucleon interpolators used in the calculation. The interpolators with asterisk (*) on them are hybrid in nature. \label{tab:interpolator} }
\end{table}

%%%%%%%%%%%%%%%%%%%%%%%%%
%%%%%%%%%%%%%%%%%%%%%%%%%
\subsection{Momentum Smearing}

To access a wide range of Ioffe-times, we perform the lattice calculation at multiple spatial momenta. On the lattice, the spatial momentum is discretized and expressed as
\be
    p = \frac{2 \, \pi \, l}{a \, L} \, .
\ee
Here, $L = 32$, is the spatial extent of the lattice. For $p$, where $l > 3$,  we enhance the overlap of the interpolators onto the lowest-lying states in motion by applying momentum smearing~\cite{Bali:2016lva}. We follow the procedure introduced in~\cite{Egerer:2020hnc} and add a phase to the distillation eigenvectors for higher momenta to preserve translational invariance, which is essential for the projection onto the states of definite momenta. The ``phased" distillation eigenvector becomes,
\be
    \Tilde{\nu}_x^{(k)} (\vv{z}, t) = e^{i \vv{\zeta} \cdot \vv{z}} \; \nu_x^{(k)} (\vv{z}, t) \, .
\ee

It is sufficient to modify the previously computed eigenvectors to perform calculation at the higher lattice momenta, though the perambulators and the elementals need to be recalculated with these ``phased" eigenvectors. For our calculation, choosing 
\bea
    \vv{\zeta} = 2 \cdot \frac{2 \pi}{L} \hat{z}
\eea
gives the momentum smearing needed for boosts up to $p = 6 \times \frac{2 \pi}{a L}$.

%%%%%%%%%%%%%%%%%%%%%%%%%
%%%%%%%%%%%%%%%%%%%%%%%%%
\section{Lattice Details}\label{sec:latt_details}
%%%%%%%%%%%%%%%%%%%%%%%%%
%%%%%%%%%%%%%%%%%%%%%%%%%

We perform our calculation on an isotropic ensemble with $(2+1)$ dynamical flavors of clover Wilson fermions with stout-link smearing~\cite{Morningstar:2003gk} of the gauge fields and a tree-level tadpole-improved Symanzik gauge action, with approximate lattice spacing, $a\sim 0.094$ fm and pion mass, $M_\pi\sim 358$ MeV, generated by the JLab/W\&M collaboration~\cite{lattices}. The rational hybrid Monte Carlo (RHMC) algorithm~\cite{Duane:1987de} is used to carry out the updates. One iteration of four-dimensional stout-smearing with the weight $\rho = 0.125$ for the staples is used in the fermion action. After stout-smearing, the tadpole-improved tree-level clover coefficient, $C_{SW}$, is very close to the non-perturbative value. This is confirmed using the Schr\"odinger functional method for determining the clover coefficient non-perturbatively~\cite{lattices}.  The tuning of the strange quark mass is done by first setting the quantity, $(2\, M^2_{K^+} - M^2_{\pi^0})/M^2_{\Omega^-}$ equal to its physical value 0.1678. This quantity is independent of the light quark masses to the lowest order in $\chi$PT, depending only on the strange quark mass~\cite{HadronSpectrum:2008xlg}. So, it can be tuned in the SU(3) symmetric limit. The resulting value of the strange quark mass is then kept fixed as the light quark masses are decreased in the (2+1) flavor theory to their physical values.

We use 64 temporal sources over 349 gauge configurations, with each configuration separated by 10 HMC trajectories. The two light quark flavors, $u$ and $d$ are taken to be degenerate and the lattice spacing was determined using the $w_0$ scale~\cite{Borsanyi:2012zs}. We summarize the parameters of the ensemble in Table~\ref{tab:latt}.

\begin{table}
  \renewcommand{\arraystretch}{1.5}
  \setlength{\tabcolsep}{10pt}
  \begin{tabular}{cccccc}
  \toprule
  ID & $a$ (fm) & $M_{\pi}$ (MeV) & $L^3 \times N_t$ & $N_{\rm cfg}$ & $N_{srcs}$\\
    \midrule
    $a094m358$ & 0.094(1) & 358(3) & $32^3 \times 64$ & 349 & 64\\
    \bottomrule
    \end{tabular}
\caption{The parameters of the ensemble used in this work. Here, $N_{\rm cfg}$ is the number of gauge configurations.\label{tab:latt}}
\end{table}

%%%%%%%%%%%%%%%%%%%%%%%%%
%%%%%%%%%%%%%%%%%%%%%%%%%
\section{Variational Analysis}\label{var_analysis}

To check whether the two-point correlators give us the expected results, we investigate the associated principal correlators and extract the energy spectra by performing a variational analysis for the nucleon at rest in the $G_{1g}$ channel and for all the boosted frames in the $\mathrm{Dic_4}$ little group with the interpolators in Table~\ref{tab:interpolator}.  This fitting procedure is discussed in detail in~\cite{Khan:2020ahz, Egerer:2018xgu, Edwards:2011jj}. We only summarize the procedure here. We solve the GEVP of Eq.~\eqref{eq:gev} over a range of $t_0$. We then define optimal interpolators, in the variational sense, for the energy eigenstates, $|n\rangle$ through $\sum_i u^i_n \mathcal{\bar{O}}_{N,i}$. Here, $\mathcal{\bar{O}}_{N,i}$ are the interpolators used in the calculation and $u^i_n$ are the weights of these interpolators that define the optimal interpolator. The energy associated with each state $|n\rangle$ is obtained by fitting its principal correlator according to
\begin{equation}
  \lambda_n(t,t_0) = (1-A_n) e^{-E_n(t-t_0)} + A_n e^{-E'_n(t-t_0)} \, . \label{eq:fit_form}
\end{equation}

In our fitting procedure, we aim to ensure that the principal correlators are dominated by the leading exponential. Thus in each of our fits, we choose $t_0$ such that we obtain an acceptable $\chi^2/{\rm d.o.f.}$, that the value of $A_n$ is small, typically less than 0.1, and that, for each principal correlator, $\lambda_n(t,t_0)$, the subleading energy $E_n'$ is larger than than the leading energies for all the principal correlators.  This indicates that the matrix of two-point correlators is to a large degree, saturated by the lowest-lying states.

\begin{figure}[!htb]
\center{\includegraphics[scale=0.22]{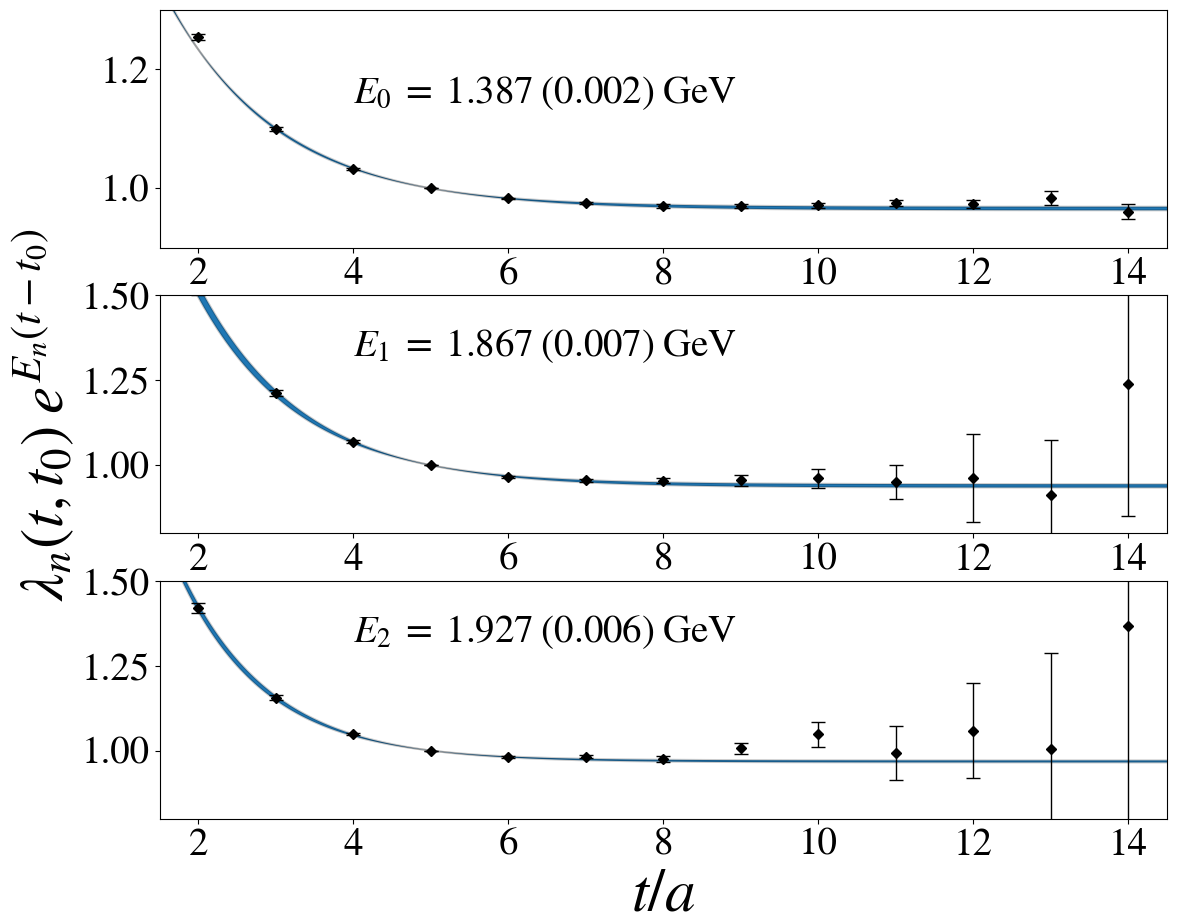}}\caption{Fits to the principal correlators for the nucleon with for $p = 2 \times \frac{2 \pi}{a L}$ =  0.82 GeV, subduced onto the little group, $\mathrm{Dic}_4$, on the ensemble $a094m358$, for $t_0 = 5$. The plots show $\lambda_n (t, t_0) \, e^{E_n (t - t_0)}$ data on the y-axes and the lattice time-slices on the x-axes; the blue bands are the two-exponential fits as described in the text. The top, middle and bottom panels show the principal correlators for the ground state, the first excited-state and the second excited-state respectively. In each panel, the energy corresponding to the leading exponential state is labelled by $E_n$. \label{fig:mom_002}}
\end{figure}

\begin{figure}[!htb]
\center{\includegraphics[scale=0.22]{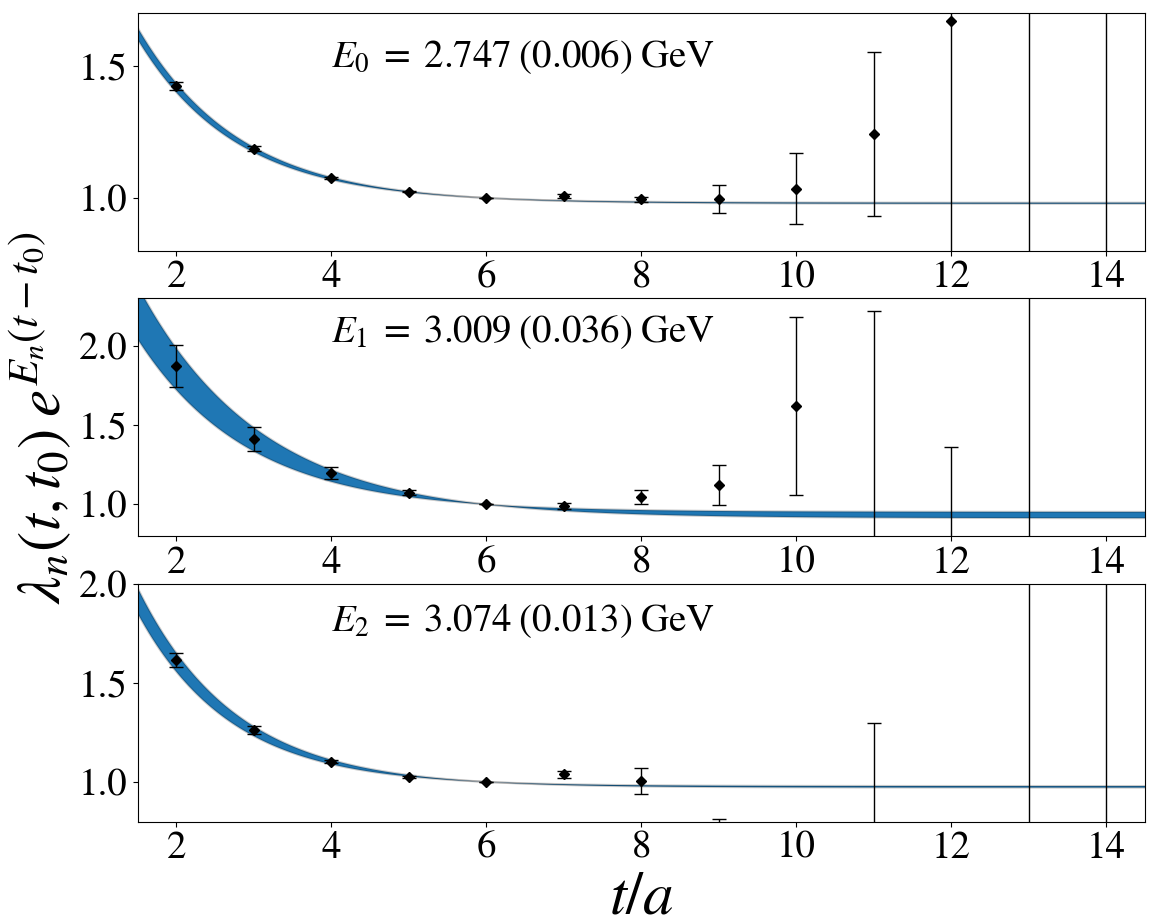}}\caption{Fits to the principal correlators for the nucleon with $p = 6 \times \frac{2 \pi}{a L}$ =  2.46 GeV, subduced onto the little group, $\mathrm{Dic}_4$, on the ensemble $a094m358$, for $t_0 = 6$. The plots show $\lambda_n (t, t_0) \, e^{E_n (t - t_0)}$ data on the y-axes and the lattice time-slices on the x-axes; the blue bands are the two-exponential fits as described in the text. The top, middle and bottom panels show the principal correlators for the ground state, the first excited-state and the second excited-state respectively. In each panel, the energy corresponding to the leading exponential state is labelled by $E_n$. \label{fig:mom_006}}
\end{figure}

In Fig.~\ref{fig:mom_002} and~\ref{fig:mom_006}, we show fits to the leading principal correlators for the nucleon subduced onto the little group, $\mathrm{Dic_4}$ for $p = 2 \times \frac{2 \pi}{a L}$ = 0.82 GeV, and $p = 6 \times \frac{2 \pi}{a L}$ =  2.46 GeV, respectively. For each panel, the blue band is the reconstruction from the fitted parameters. The approach of the plateaux close to unity at large times is indicative of the small value of $A_n$ in the fits, and the small contribution of the other states to each principal correlator.

\begin{figure}[!htb]
\center{\includegraphics[scale=0.7]{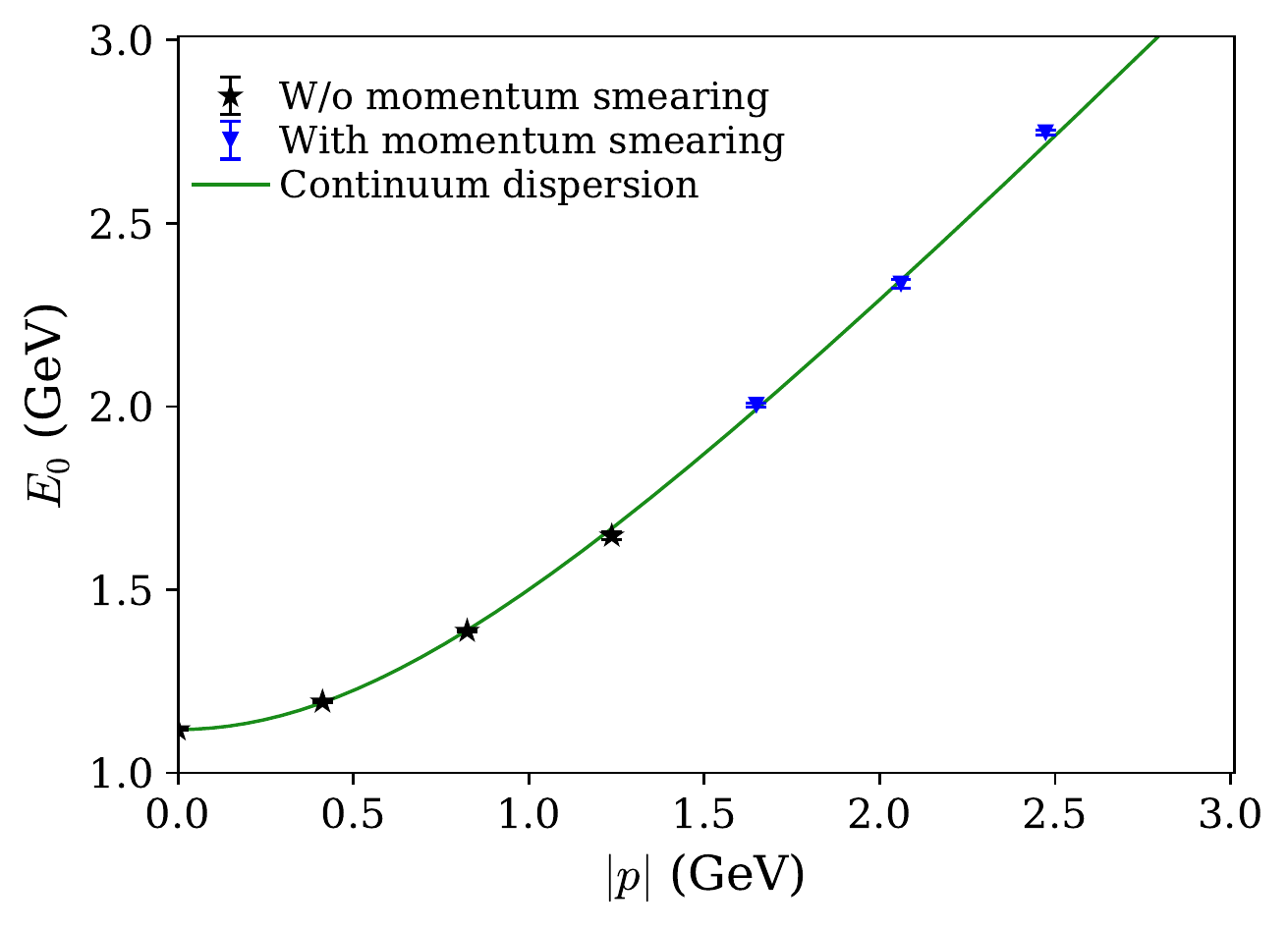}}\caption{The ground state nucleon dispersion relation on the ensemble $a094m358$, the solid line being the continuum dispersion relation. Energies without phasing are in black and energies with phasing are in blue.
\label{fig:disp_plot}}
\end{figure}

In Fig.~\ref{fig:disp_plot}, we plot the ground state nucleon energies extracted using the variational analysis with respect to the spatial momentum, together with the expectations from the continuum dispersion relation.

Fig.~\ref{fig:disp_plot} shows that for lower momenta, the unphased ground state nucleon energies are in excellent agreement with the continuum dispersion relation. At $p = 3 \times \frac{2 \pi}{a L}$ = 1.23 GeV, the ground state energy starts to deviate, but from $p = 4 \times \frac{2 \pi}{a L}$ =  1.64 GeV, after phasing, the ground state energy starts to align with the continuum dispersion curve, indicating that adding a phase to the distillation eigenvectors with $\zeta = 2 \frac{2 \pi}{L}$ resulted in a significant increase in the overlap of the interpolators onto the lowest-lying states in motion.

%%%%%%%%%%%%%%%%%%%%%%%%%
%%%%%%%%%%%%%%%%%%%%%%%%%
\section{Matrix Element Extraction}\label{sec:mtx_calc}
%%%%%%%%%%%%%%%%%%%%%%%%%

\subsection{Three-point Correlator}

We calculate the matrix elements by first computing the three-point correlators by inserting gluonic currents between the source and the sink of the two-point correlators. The three-point correlator can be expressed as
\be
\langle C_{3pt}(t, t_g) \rangle = \langle 0 | \mathcal{T} \{ \mathcal{O}_N(t) \, O_g(t_g) \, \mathcal{\bar{O}}_N (0) \} | 0 \rangle \, ,
\ee
where $\mathcal{\bar{O}}_N$ and $\mathcal{O}_N$ are the interpolators and $t$ is the source-sink separation. The $\langle \dots \rangle$ indicates the ensemble average and $\mathcal{T} \{ \dots \}$ stands for the time-ordered product. The three-point correlator can be rewritten as
\bea
    C^i_{3pt} (t, t_g) = \Big( C^i_{2pt}(t) - \bigl\langle C_{2pt}(t) \bigr\rangle \Big) 
 \Big( O^i_g(t_g) - \langle O_g(t_g) \rangle \Big) \, ,
\eea
where $ C_{2pt}(t)$ is the nucleon two-point correlator with source-sink separation $t$ in lattice units and $t_g$ is the time-slice on which the gluonic current is inserted.

%%%%%%%%%%%%%%%%%%%%%%%%%
%%%%%%%%%%%%%%%%%%%%%%%%%
\subsection{sGEVP Method}

We implement the sGEVP method~\cite{Bulava:2011yz, Blossier:2009kd} to extract the matrix elements from the three-point correlators, a combination of the summation method~\cite{Bouchard:2016heu} and GEVP~\cite{Edwards:2011jj} method which begins with the formation of the summed three-point correlation functions formed from our basis of interpolating operators
\be
C^{i,s}_{3pt}(t) = \sum_{t_g = 1}^{t-1} C^i_{\rm 3pt}(t, t_g).
\ee

We provide details of the method in appendix~\ref{sgevp:appendix}, but the salient feature is that
for sGEVP, the systematic error decays as $ \big[ \, t \; \mathrm{exp}(-\Delta E \, t) \big]$, which is much faster than the $ \big[ \, \mathrm{exp}(-\Delta E^\prime \, t) \big]$ decay for GEVP~\cite{Edwards:2011jj}. This allows us to access the matrix elements at a much smaller temporal separation than would be possible with GEVP. This is crucial for hadron structure calculations, since the signals tend to be heavily contaminated by noise as the temporal separation is increased. sGEVP utilizes the lowest-lying spectra, conveniently calculated using distillation, by rotating the three-point correlator matrix by a suitable angle, removing much of the excited-state contaminations, and therefore performs better than the summation method~\cite{Bouchard:2016heu}, which involves only the ground-state nucleon.

In principle, increasing the number of states, $N$, in the sGEVP analysis should lead to a larger $\Delta E$, which enables matrix elements to be extracted from even smaller temporal separations. This, however, also increases the computational cost, because the $N \times N$ correlator matrix needs to be constructed, and makes solving the GEVP for the nucleon two-point correlator matrix more challenging.

%%%%%%%%%%%%%%%%%%%%%%%%%
%%%%%%%%%%%%%%%%%%%%%%%%%
\subsection{Bare Matrix Elements}

Our calculation requires the extraction of the matrix elements at multiple flow times, multiple nucleon momenta and multiple separations between the gluon fields. We perform the calculation for flow times $\tau/a^2$ = 1.0, 1.4, 1.8, 2.2, 2.6, 3.0, 3.4 and 3.8. For each flow time, we calculate the matrix elements for nucleon momenta, $p = \frac{2 \pi l}{a L}$ where $l$ = 0 to 6, and for field separations, $z = s \, a$ where $s$ = 0 to 6; $a$ being the lattice spacing. We construct the effective matrix element, $\mathcal{M}^{\rm eff}(t, z, p, \tau)$ for each flow time, nucleon momentum and field separation, using the formulation described in appendix~\ref{sgevp:appendix} and fit the matrix elements using the functional form in Eq.~\eqref{sGEVP_fit}, which can be written in simplified notation and arguments as
\bea \label{fiteq}
\mathcal{M}^{\rm eff}(t) = A + B\,t \exp(-\Delta E\, t) \, .
\eea
Here, $A$ is the matrix element we wish to extract. To perform the fit of Eq.~\eqref{fiteq} for a particular nucleon momentum, $p$, we first fit the matrix element for $z = 0$ using a Bayesian analysis and determine the corresponding fitted value of the parameter, $\Delta E$. As the hadronic spectrum is determined by the two-point correlators, we use the value of $\Delta E$ obtained from the fit to the matrix element for $z = 0$ as the prior for our subsequent fits to the matrix elements for $z > 0$ at that particular nucleon momentum. We set the prior-width of $\Delta E$ for $z > 0$ to be three times larger than the uncertainty in $\Delta E$ and allow for random priors in XMBF~\cite{XMBF}. The priors are chosen randomly from normal distributions with the given prior-widths. We perform a simultaneous and correlated fit to the matrix elements for $z = \{1, \, 2, \, 3, \, 4, \, 5, \, 6\} \times a$ = 0.094 fm, 0.188 fm, 0.282 fm, 0.376 fm, 0.470 fm, 0.564 fm respectively,
\bea \label{fiteqsimul}
\mathcal{M}^{\rm eff}(t)_i = A_i+ B_i\,t \exp(-\Delta E\, t) \, ,
\eea
where $i=1, 2, \cdots \, 6$ and the $\Delta E$ is assumed to be the same for matrix elements at a fixed nucleon momentum and flow time. This procedure is particularly helpful for a well-controlled fit to the large momentum matrix elements for which the signal-to-noise ratio is poor, especially at flow times $\tau/a^2< 1.6$.

In Fig.~\ref{fig:Mefftau}, we illustrate our fits to the matrix elements for $\tau/a^2=1.0$, in the upper row and for $\tau/a^2=3.0$ in the bottom row. Here, we compare the fitted matrix elements among the momenta, $p = \{ 1, \, 6\} \times \frac{2 \pi}{a L}$ = 0.41 GeV, 2.46 GeV respectively; and the separations, $z = \{ 0, \, 1, \, 6\} \times a $ = 0, 0.094 fm, 0.564 fm respectively,  and list the fitted parameters in Table~\ref{tab:fitparams}.  One can immediately see that the $\Delta E$ values determined for the non-zero separations are almost identical compared to that obtained for the matrix elements at $z=0$ where no prior is assigned on the fit parameter $\Delta E$. This, along with the goodness of the fit in the extraction of the matrix elements for the non-zero separations, indicates the validity of our fitting procedure.

%%%%%%%%%%%%%%%%%%%%%%%%%%%%%%%
\befs 
\centering

\includegraphics[scale=0.4]{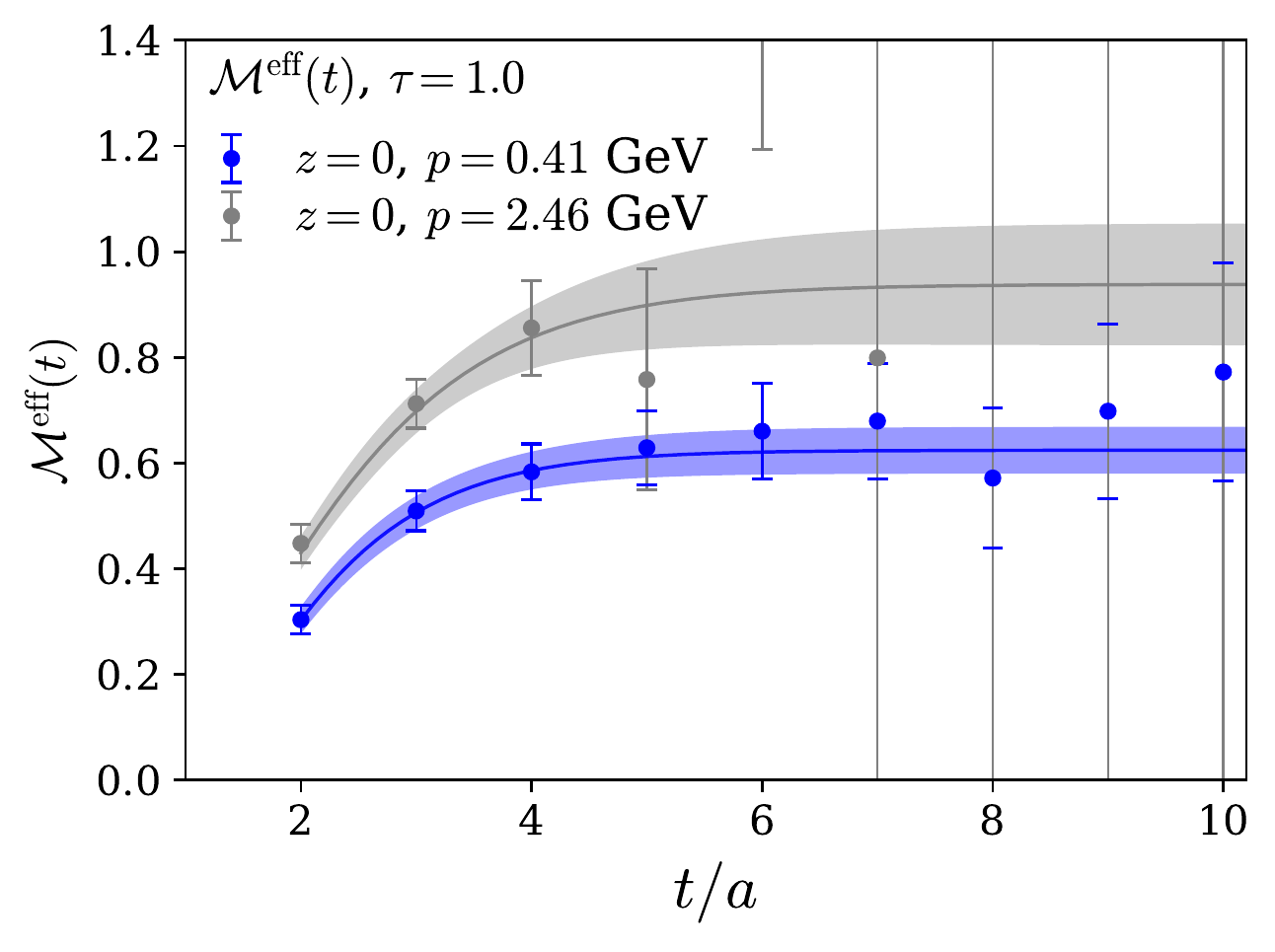}
\includegraphics[scale=0.4]{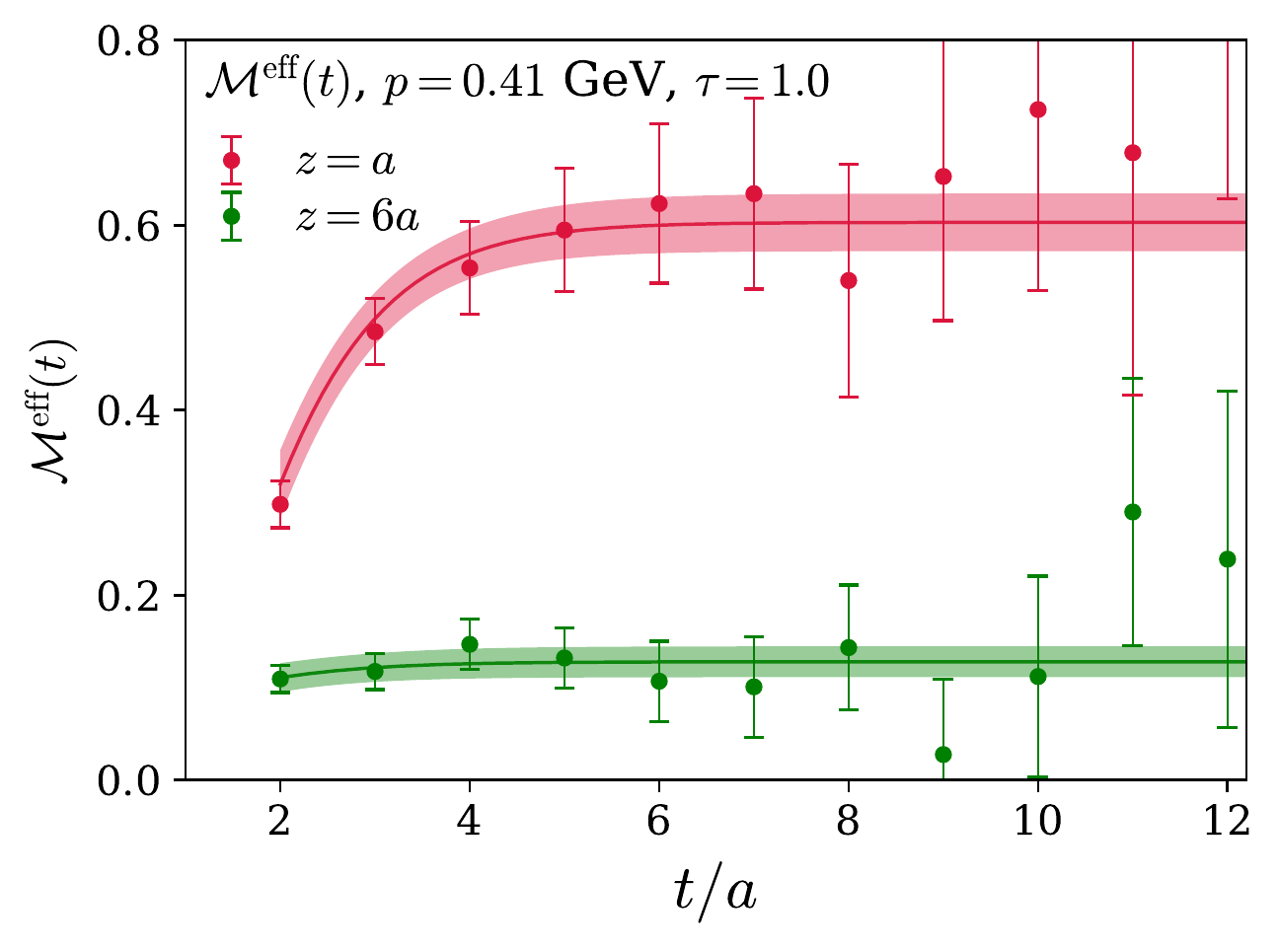}
\includegraphics[scale=0.4]{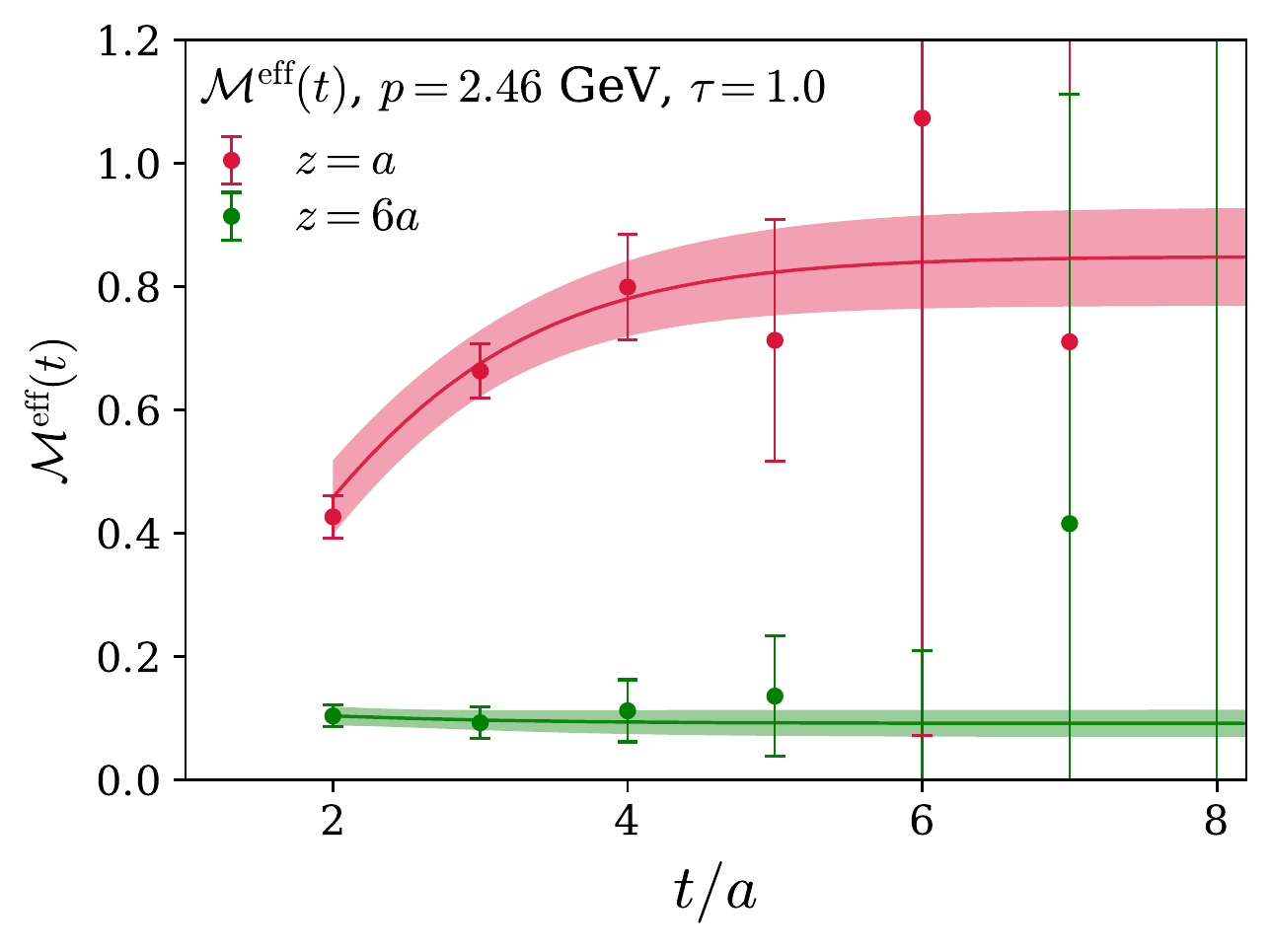}

\includegraphics[scale=0.4]{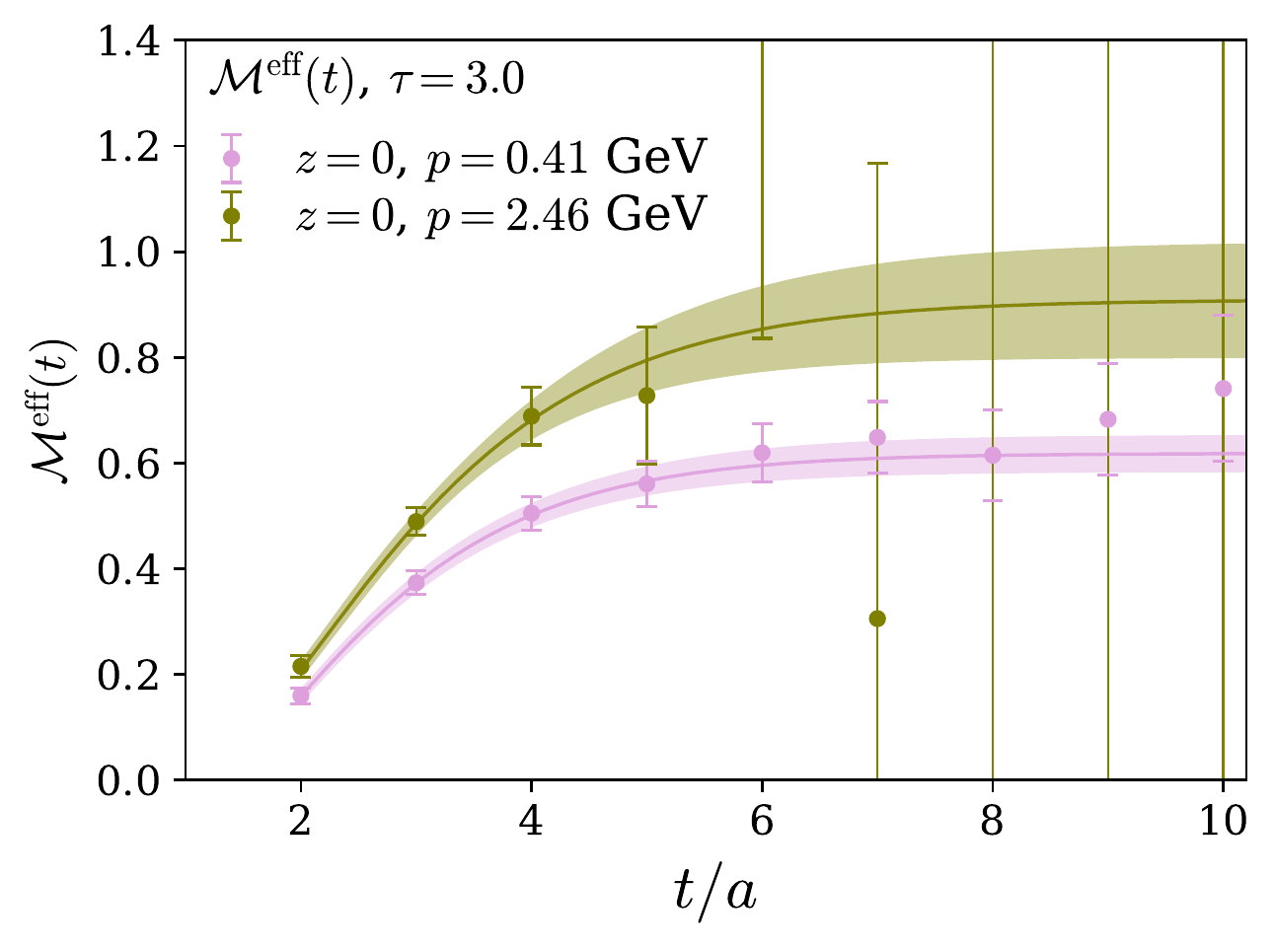}
\includegraphics[scale=0.4]{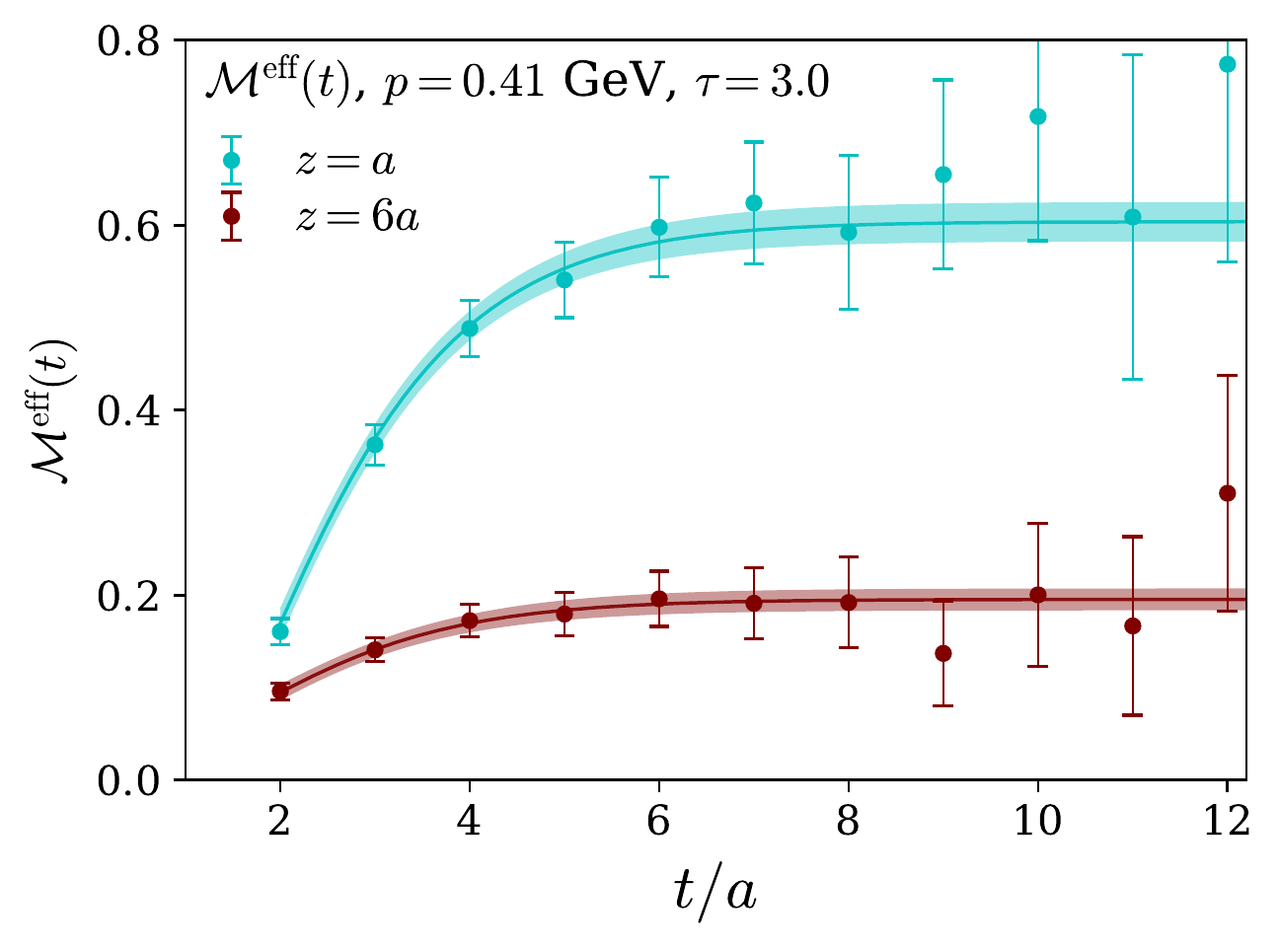}
\includegraphics[scale=0.4]{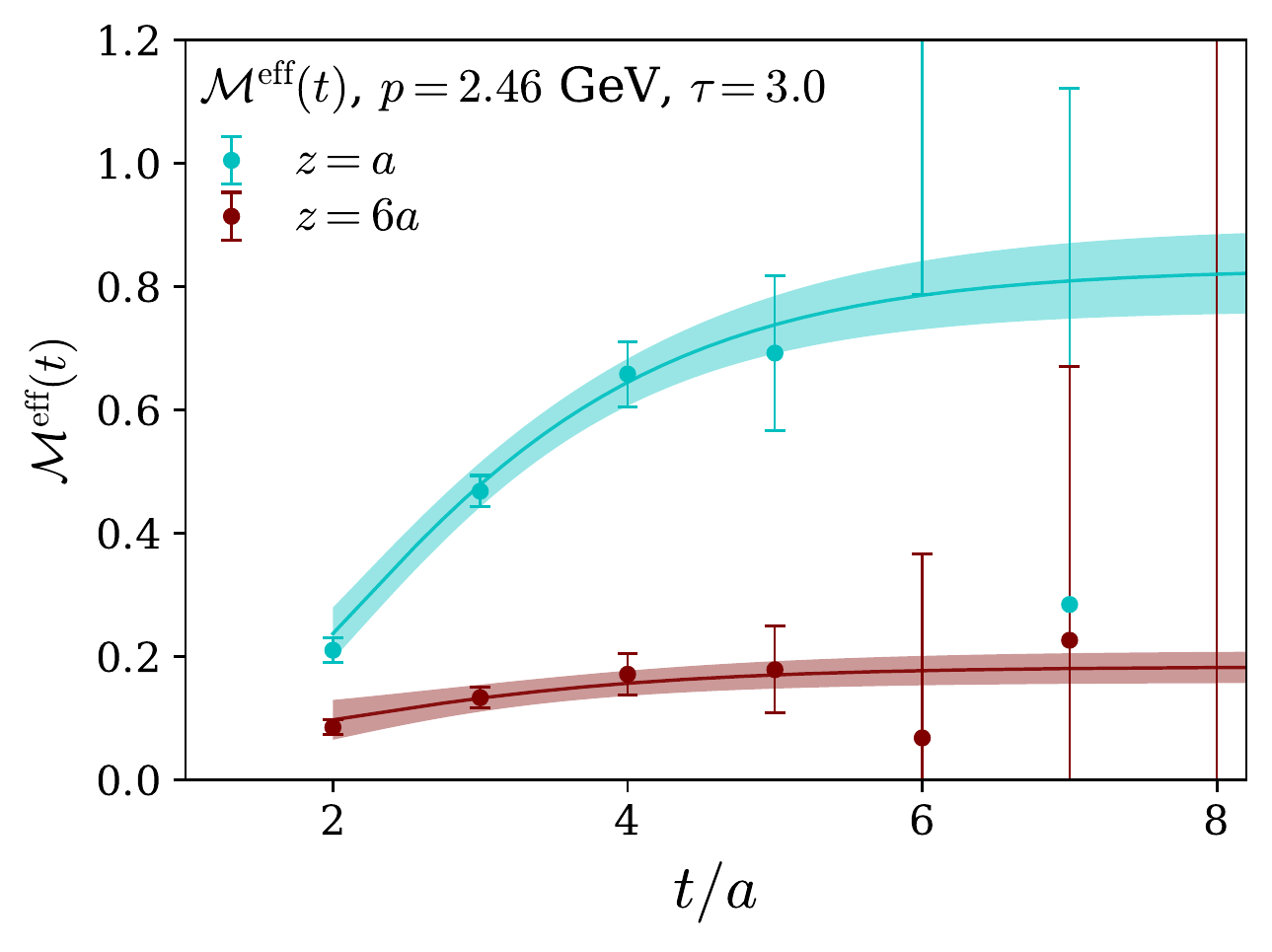}

\caption{\label{fig:Mefftau} 
Extraction of the matrix elements using the sGEVP method for different flow times, nucleon momenta and field separations on the ensemble $a094m358$. The bands are the fits described in the text. The top and bottom rows contain the matrix elements for flow time $\tau/a^2$ = 1.0 and 3.0, respectively. In each row, the left column compares between the matrix elements for $p = \{ 1, \, 6\} \times \frac{2 \pi}{a L}$ = 0.41 GeV, 2.46 GeV respectively at zero separation; the middle column compares between the matrix elements for $p = 1 \times \frac{2 \pi}{a L}$ = 0.41 GeV and separations $z = \{1, \, 6\} \times a $= 0.094 fm, 0.564 fm respectively. The right column does the same comparison as done in the middle column, but for $p = 6 \times \frac{2 \pi}{a L}$ = 2.46 GeV.
}      
\eefs{mockdemocn}
%%%%%%%%%%%%%%%%%%%%%%%%%%%%%%%

%%%%%%%%%%%%%%%%%%%%%%%
\begin{table*}
  \centering
  \setlength{\tabcolsep}{10pt}
  \renewcommand{\arraystretch}{1.5}
  \begin{tabular}{ccccccccc}
  \toprule
    $\tau/a^2$ &  $p \,$(GeV) & $z \,(a)$ & $\nu$ &  $A$ & $B$ & $\Delta E$ & $\chi^2/{\rm d.o.f.}$ \\
    \midrule
    $1.0$ & $0.41$ & $0$ & 0.00 & 0.62(4) & -2.69(79) & 1.41(18) & $0.53$ \\
    $1.0$ & $0.41$ & $1$ & 0.20 & 0.60(3) & -2.35(50) & 1.40(13) & $0.77$ \\
    $1.0$ & $0.41$ & $6$ & 1.18 & 0.13(2) & -0.14(7) & 1.40(13) & $0.77$ \\
    $1.0$ & $2.46$ & $0$ & 0.00 & 0.94(12) & -2.56(83) & 1.15(25) & $0.62$ \\
    $1.0$ & $2.46$ & $1$ & 1.18 & 0.85(8) & -2.23(28) & 1.22(12) & $0.29$ \\
    $1.0$ & $2.46$ & $6$ & 7.07 & 0.09(2) & 0.07(13) & 1.22(12) & $0.29$ \\
    $3.0$ & $0.41$ & $0$ & 0.00 & 0.62(4) & -1.80(13) & 1.03(5) & $0.35$ \\
    $3.0$ & $0.41$ & $1$ & 0.20 & 0.60(2) & -1.68(8) & 1.02(4) & $0.31$ \\
    $3.0$ & $0.41$ & $6$ & 1.18 & 0.19(1) & -0.39(4) & 1.02(4) & $0.31$ \\
    $3.0$ & $2.46$ & $0$ & 0.00 & 0.91(11) & -2.16(20) & 0.91(10) & $0.29$ \\
    $3.0$ & $2.46$ & $1$ & 1.18 & 0.83(7) & -1.90(17) & 0.93(7) & $0.22$ \\
    $3.0$ & $2.46$ & $6$ & 7.07 & 0.18(3) & -0.28(13) & 0.93(7) & $0.22$ \\
    \bottomrule
  \end{tabular}
\caption{The fitted parameters and the goodness of the fits for the matrix elements shown in Fig.~\ref{fig:Mefftau}. For a particular flow time and nucleon momentum, we first fit the matrix elements at $z=0$; the information regarding the fit parameter $\Delta E$ from this fit is used to set the prior for $\Delta E$ in a simultaneous correlated fit for the matrix elements of all the non-zero separations.}\label{tab:fitparams}
\end{table*}

From Fig.~\ref{fig:Mefftau} and the corresponding fit parameters in Table~\ref{tab:fitparams} we see that the lattice data are described well by our fit procedure. The $\chi^2/{\rm d.o.f.}$ shows that the choice of prior-width for $\Delta E$ at $z > 0$ is an appropriate one. We notice from Fig.~\ref{fig:Mefftau} that the matrix elements for $z = 6a $ = 0.564 fm, have a flat behavior as a function of the source-sink separations. This can also be understood from the smallness of $B$-parameters listed in Table~\ref{tab:fitparams}, with relatively larger uncertainties. 

The nucleon two-point correlators have quite good signal-to-noise ratios up to the source-sink separation $t = 9a$ = 0.846 fm at $p = 6 \times \frac{2 \pi}{a L}$ = 2.46 GeV, as can be seen from Fig.~\ref{fig:mom_006}. Fig.~\ref{fig:Mefftau} shows, however, that the matrix elements almost lose any statistical signal around source-sink separation $t = 6a$ = 0.564 fm, which is expected as the nucleon momentum increases. As shown in~\cite{Dudek:2012gj}, the optimized interpolators reduce the excited-state contributions allowing us to start the fit at significantly earlier source-sink separations. In support of this, we indeed see from  Fig.~\ref{fig:Mefftau} that the matrix elements for $p = 1 \times \frac{2 \pi}{a L}$ = 0.41 GeV reach a plateau around the source-sink separation, $t = 4a$ = 0.376 fm.

We note that lattice QCD calculations of the gluonic observables are, in general, much noisier than quark matrix elements. Measures of the  goodness of the fits do not necessarily reflect all the systematic uncertainties in our extractions of the fit parameters $A$, $B$, and $\Delta E$. However, by using  $N$ interpolators within a variational approach, we are better able to sample the Hilbert space in a particular irrep.\ in finite volume. This has been proven successful in nucleon structure calculation in~\cite{Egerer:2018xgu}. The crucial insight is that projecting to the definite finite volume states via the variational solutions allows us to take advantage of the orthogonality of the states in the Hilbert space~\cite{Blossier:2009kd}. There are clearly residual excited-states present as constructing the ideal basis is unrealistic. However, a significant improvement is achieved by incorporating a moderate number of interpolators and applying distillation, one of the most computationally cost-effective methods for implementing a large number of interpolators. Therefore, by adding multiple interpolators we have attempted to systematically improve the determination of $A$, $B$, and $\Delta E$ in this calculation. Further investigation with larger statistics will be necessary for complete estimate of all the systematic uncertainties associated with excited-state contamination at large nucleon momenta.

%%%%%%%%%%%%%%%%%%%%%
%%%%%%%%%%%%%%%%%%%%%
\befs 
\centering

\includegraphics[scale=0.6]{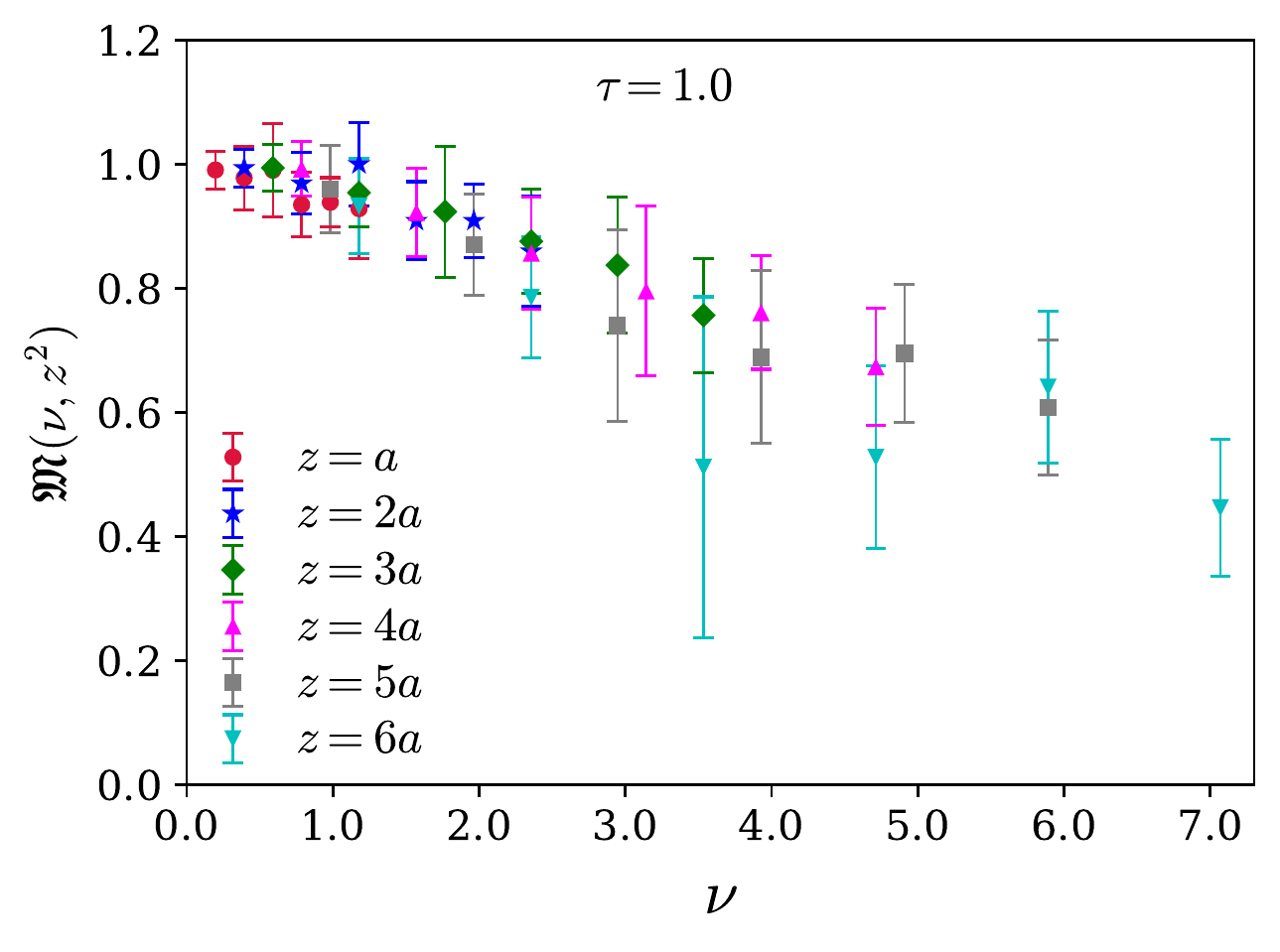}
\includegraphics[scale=0.6]{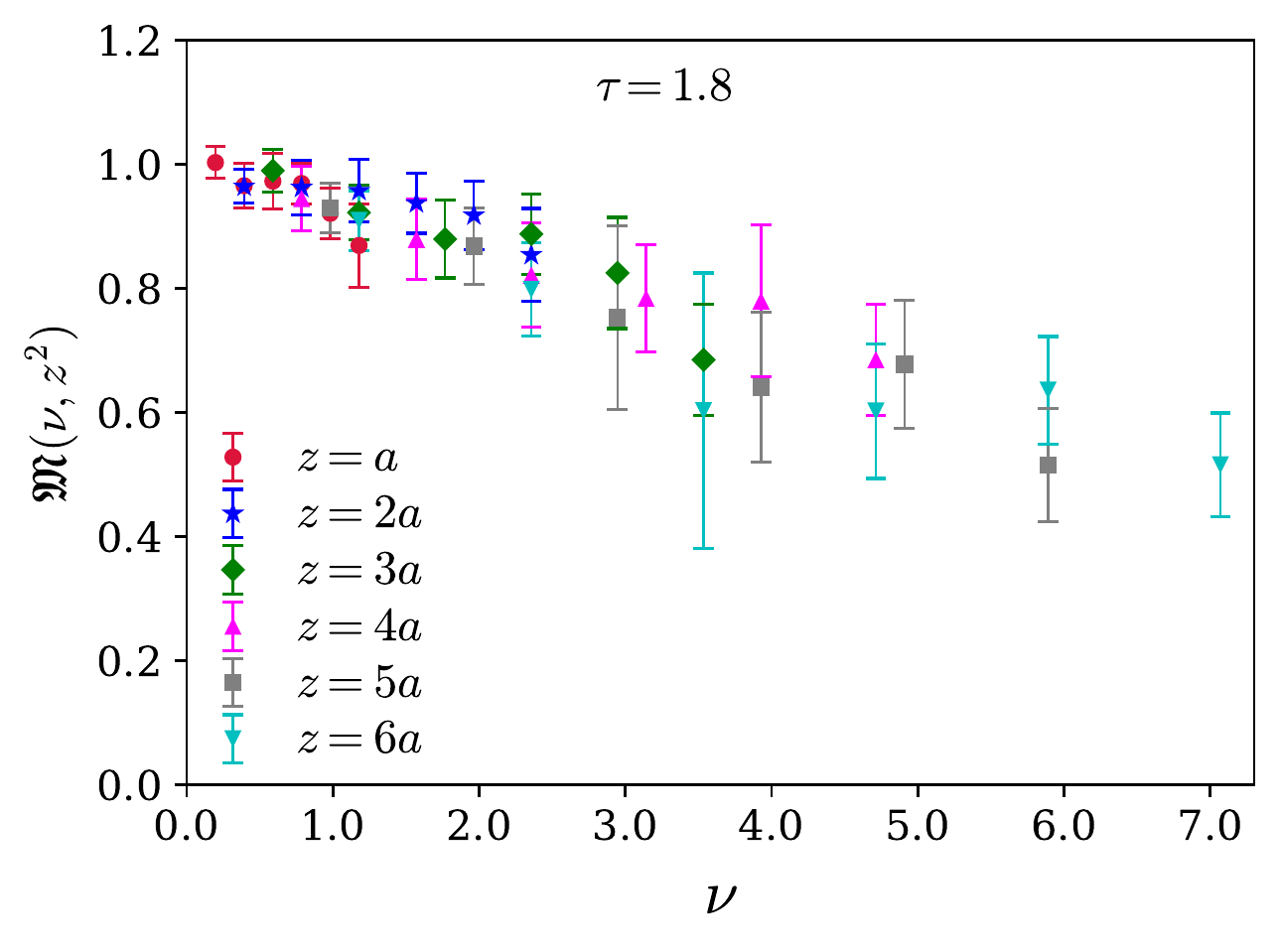}

\includegraphics[scale=0.6]{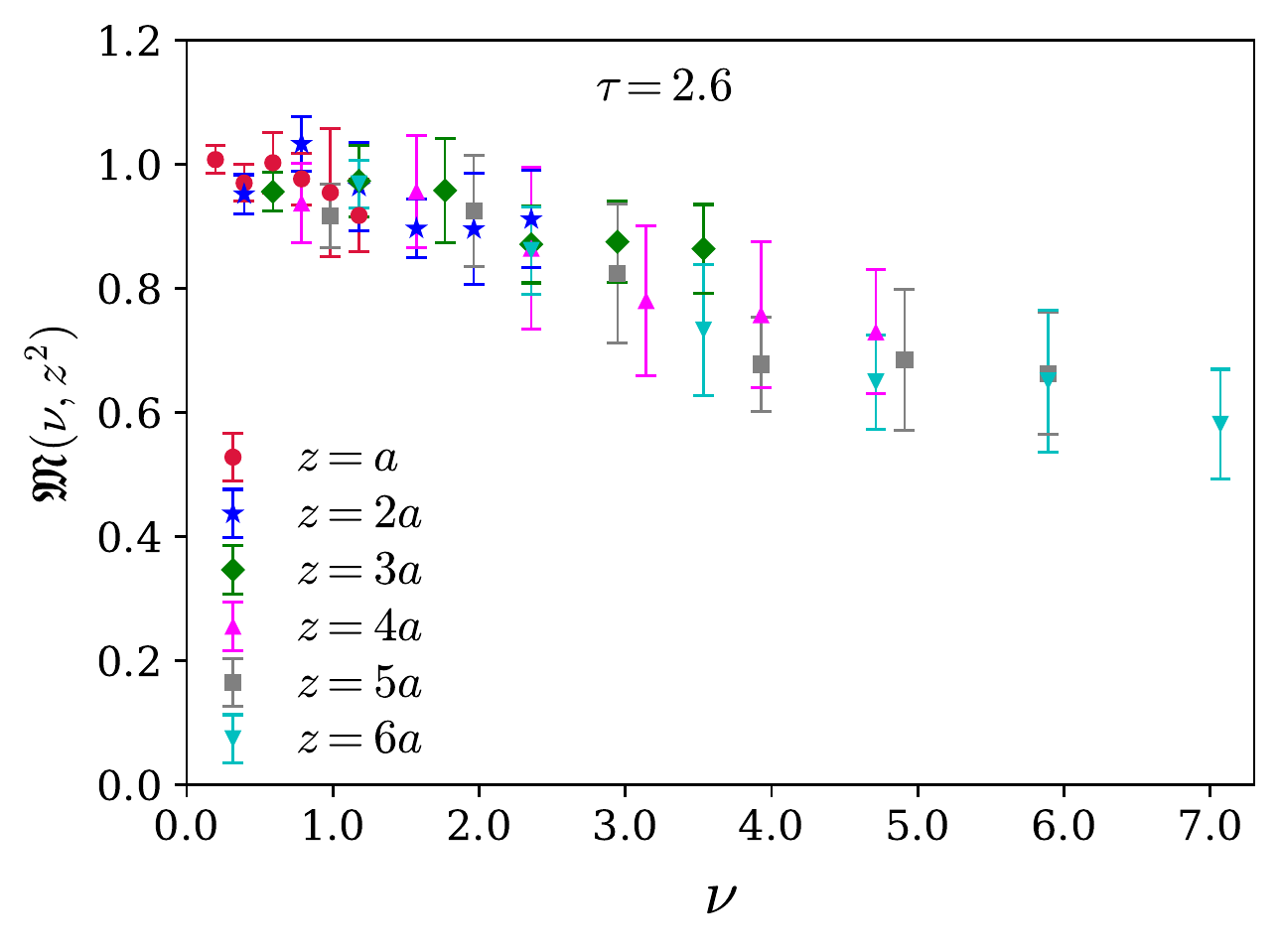}
\includegraphics[scale=0.6]{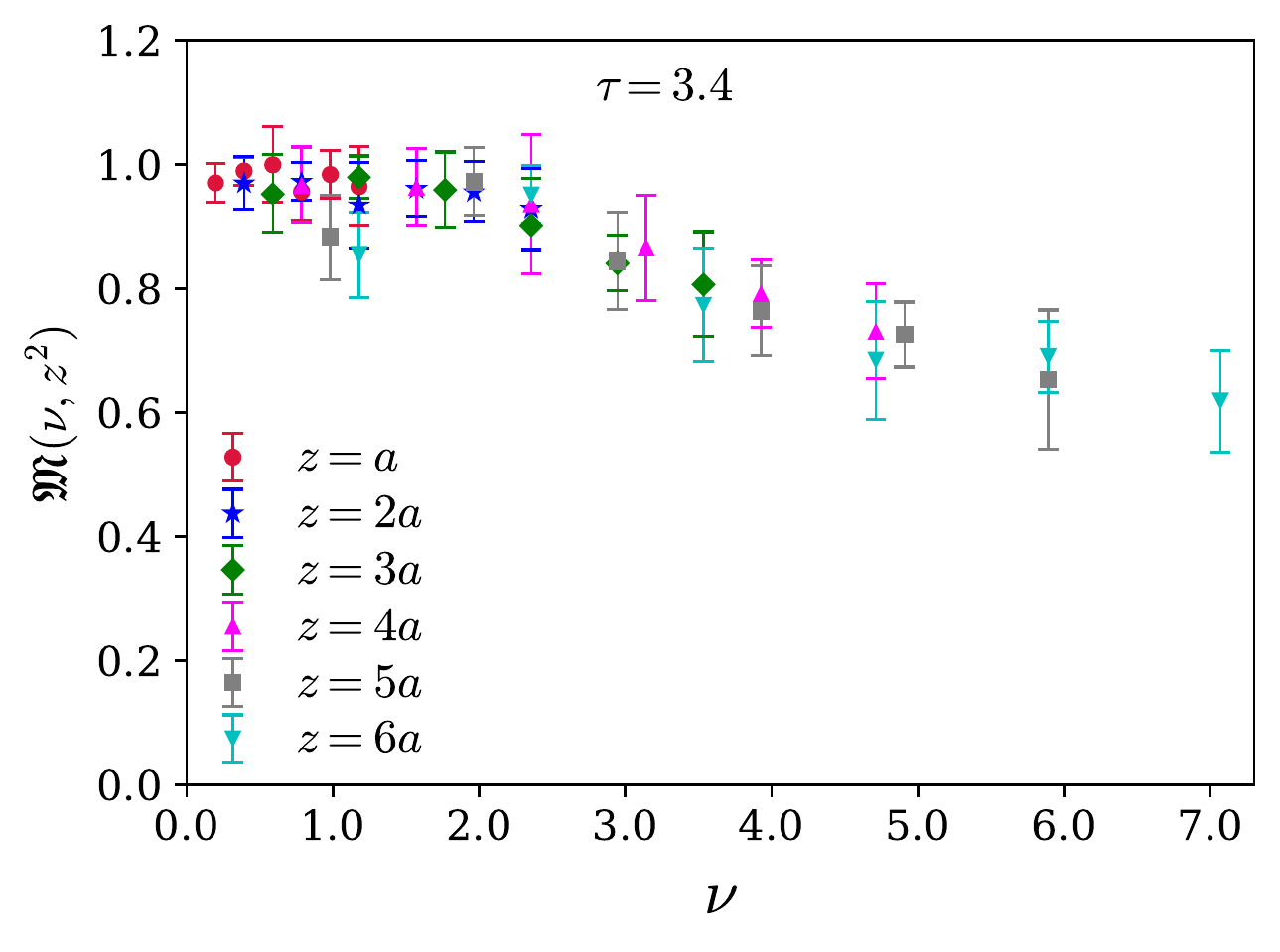}

\caption{\label{fig:ritdall} The reduced matrix elements, $\mathfrak{M} (\nu, z^2)$ with respect to the Ioffe-time for different flow times. The top-left, top-right, bottom-left, bottom-right panels have the reduced matrix elements for $\tau$ = 1.0, 1.8, 2.6, 3.4 in lattice units respectively.}
\eefs{mockdemocn}

%%%%%%%%%%%%%%%%%%%%%%%%%%%%
%%%%%%%%%%%%%%%%%%%%%%%%%%%%
\subsection{Reduced Matrix Elements and Zero Flow time Extrapolation}\label{sec:zero-flowtime}

From the bare matrix elements, we calculate the reduced matrix elements using the double ratio in Eq.~\eqref{eq:doubleratio} for different flow times, nucleon momenta and field separations. We present the reduced matrix elements for four different values of $\tau/a^2$ in Fig.~\ref{fig:ritdall}. 
We expect the higher twist contributions, discretization effects, and flow time dependence to be minimized through this double ratio.

\begin{figure}[!htb]
\center{\includegraphics[scale=0.7]{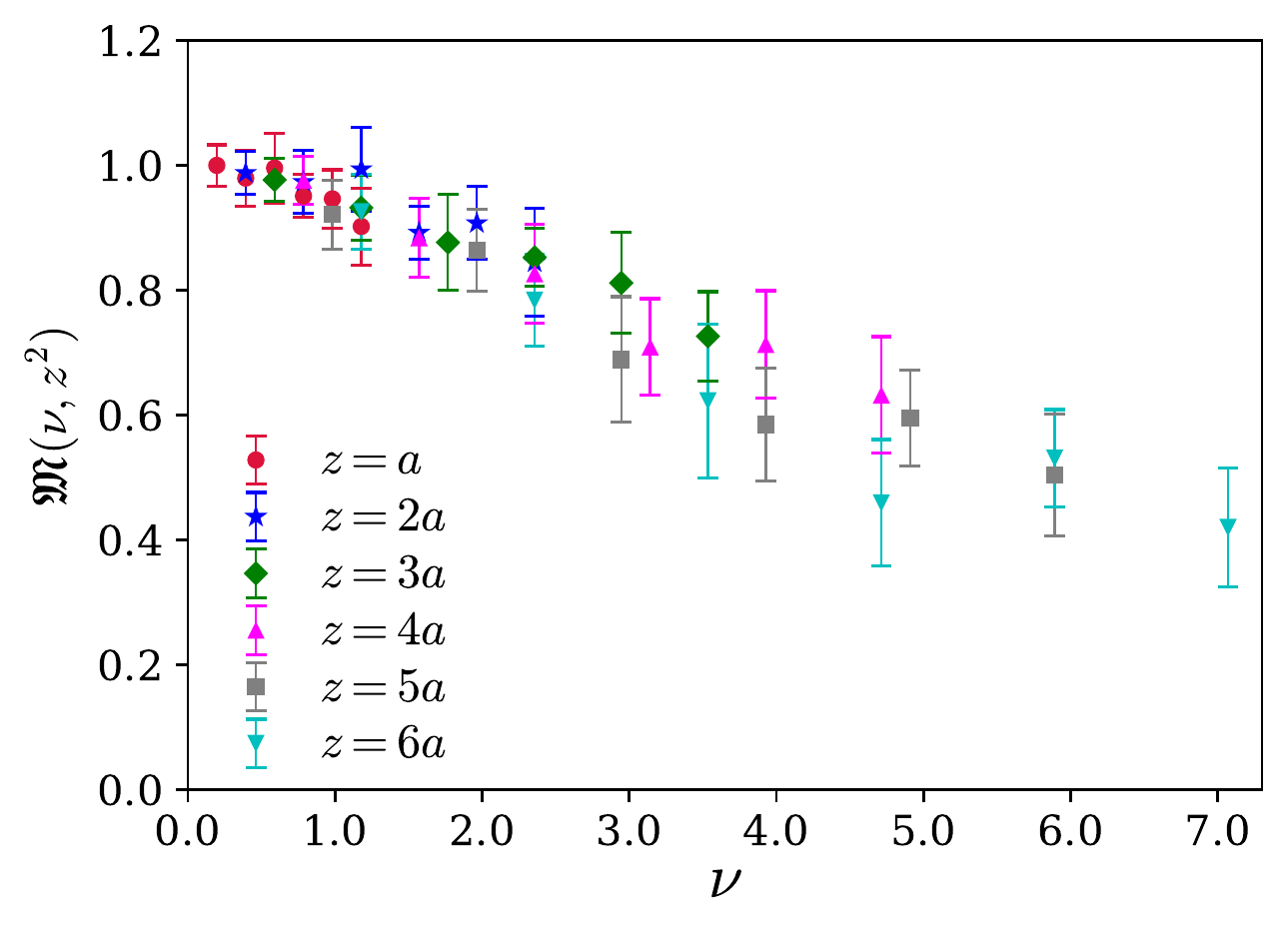}}
\caption{\label{fig:pseudo-rITDfinal} Reduced Ioffe-time pseudo-distribution, $\mathfrak{M} (\nu, z^2)$ plotted with respect to the Ioffe-time $\nu$. For each nucleon momentum and field separation, the reduced matrix elements for different flow times are extrapolated to the limit, $\tau \to 0 \,$, extracting the flow time independent reduced pseudo-ITD.} 
\end{figure}

From the reduced matrix elements at different flow times, we calculate the reduced pseudo-ITD distribution by extrapolating to zero flow time. At fixed values of the field separation, $z$, and nucleon momentum, $p$, we find that the $\tau$-dependence is best fit by a linear form, $\mathfrak{M}(\tau) = c_0 + c_1 \tau$, which we use to determine the reduced pseudo-ITD matrix elements for the subsequent analyses. The values of the fitted parameters are tabulated in appendix~\ref{zero_flow_time_reduced_mtx_elem}. Out of 36 different fits, we present six examples of such extrapolation in Fig.~\ref{fig:tauextrapolation} and for all extrapolations, we find $\chi^2/{\rm d.o.f.}< 1.0$.  Finally, we present the reduced pseudo-ITD in the zero flow time limit in Fig.~\ref{fig:pseudo-rITDfinal}.

%%%%%
\befs 
\centering

\includegraphics[scale=0.4]{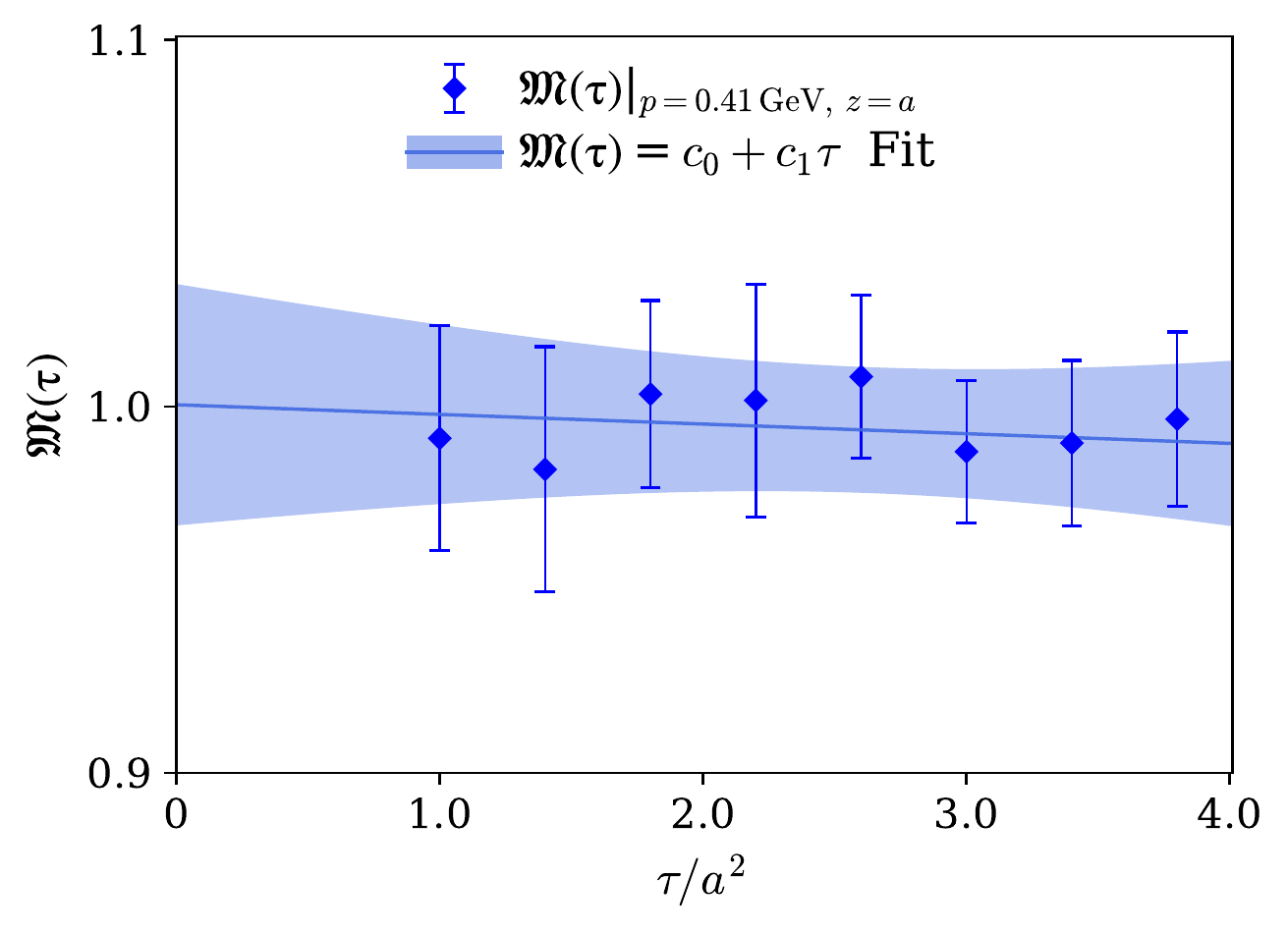}
\includegraphics[scale=0.4]{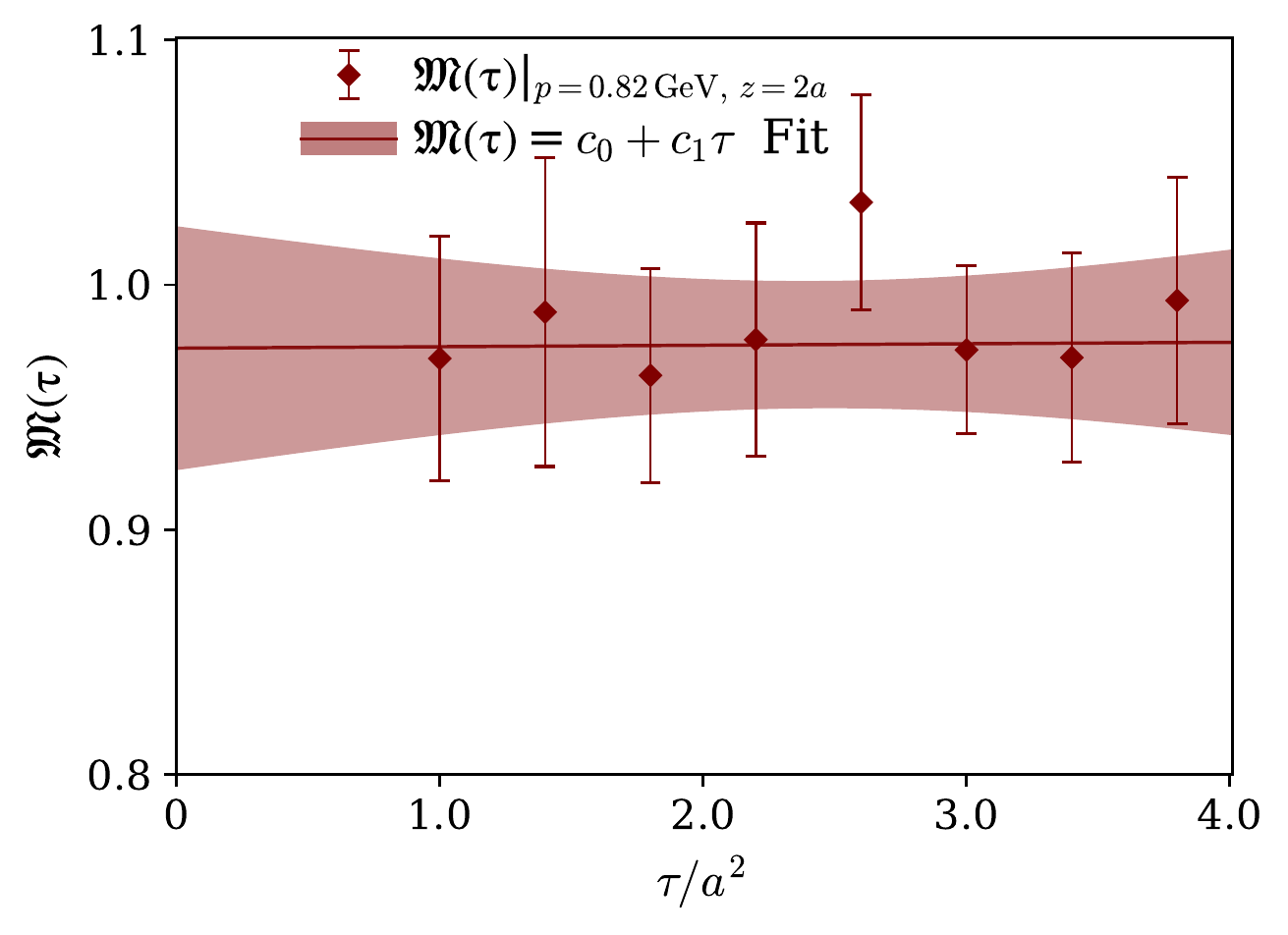}
\includegraphics[scale=0.4]{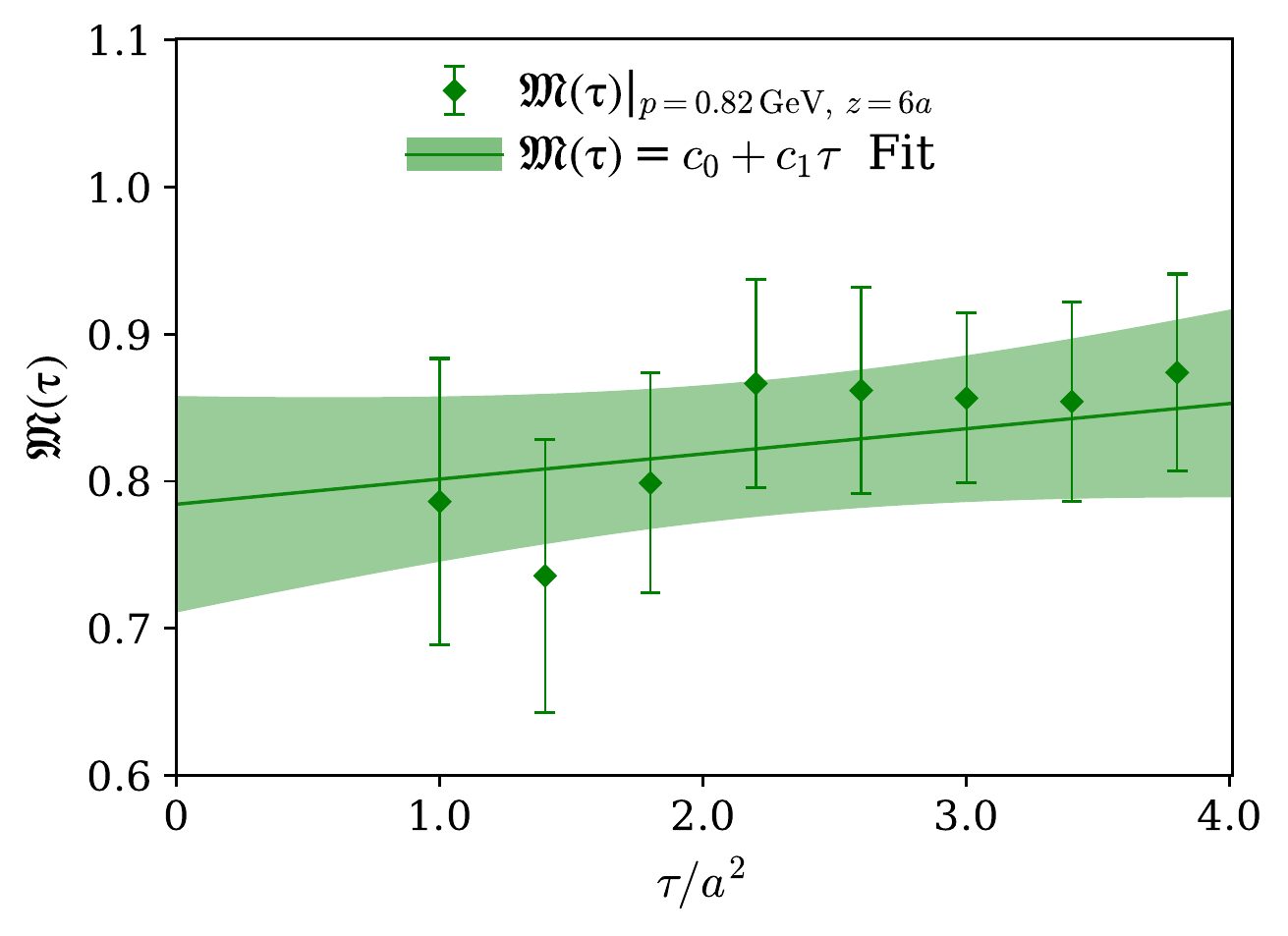}

\includegraphics[scale=0.4]{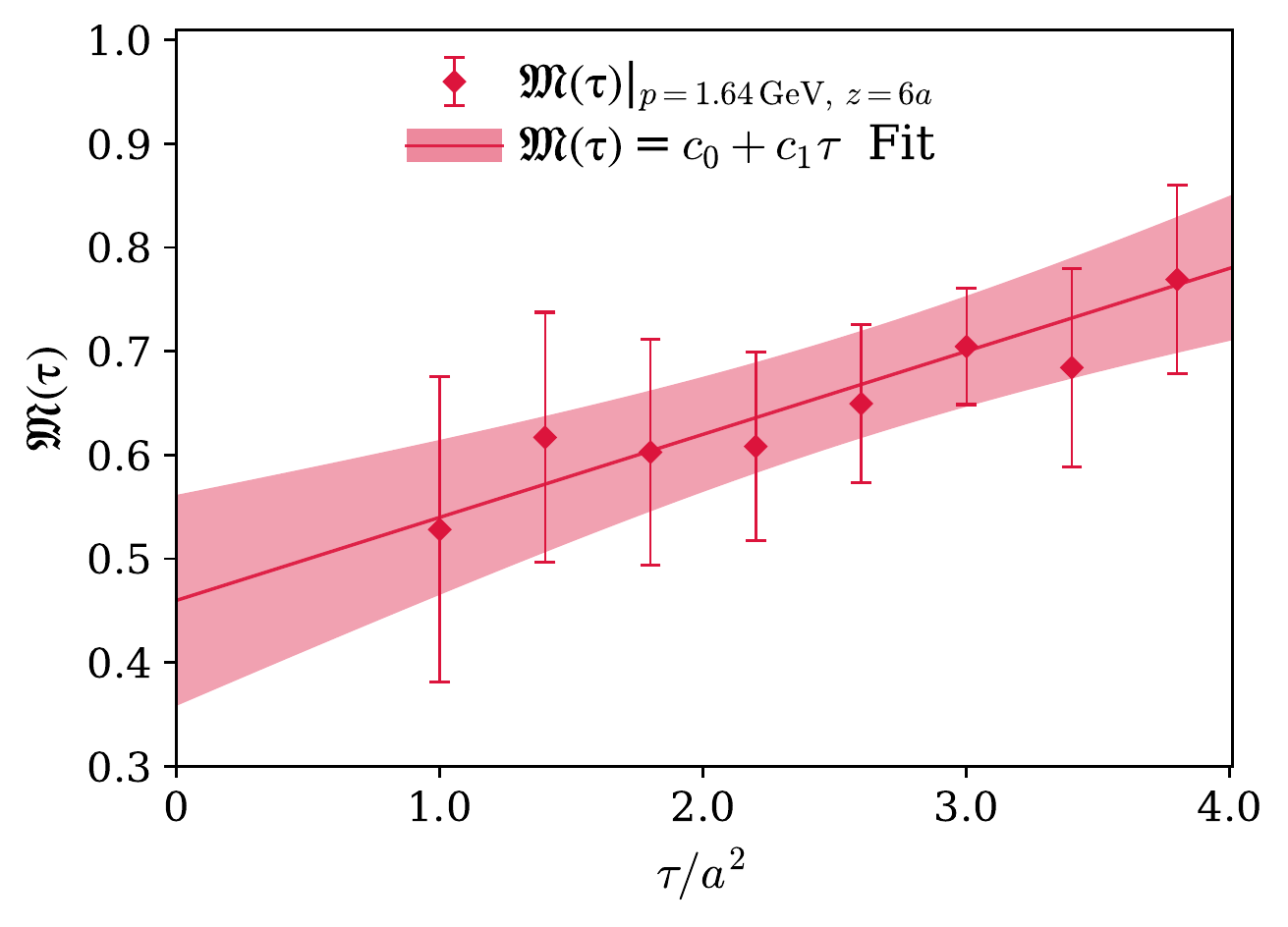}
\includegraphics[scale=0.4]{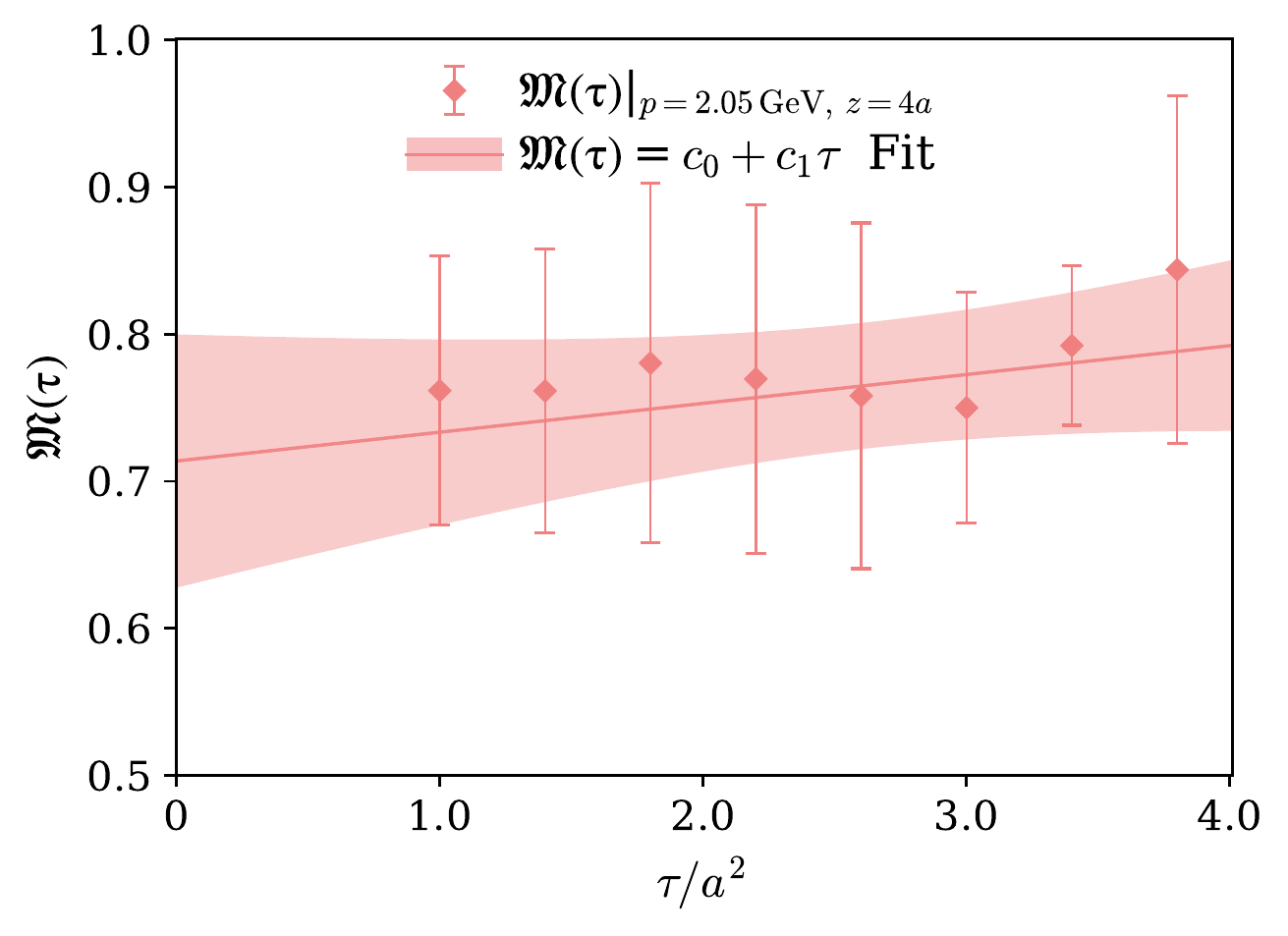}
\includegraphics[scale=0.4]{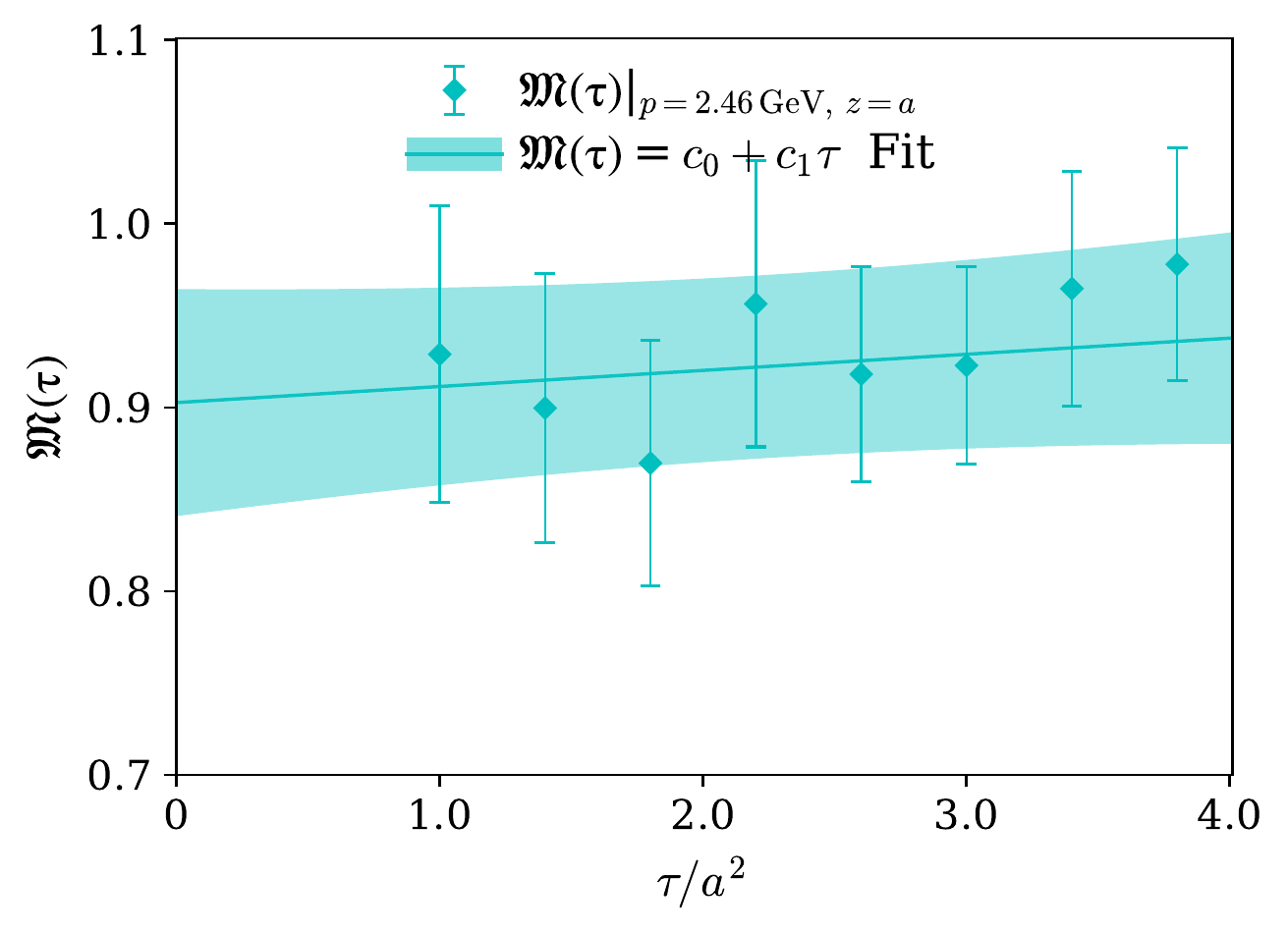}

\caption{\label{fig:tauextrapolation} Reduced matrix elements, $\mathfrak{M} (\tau)$ extrapolated to $\tau \to 0$ limit for different nucleon momenta and different field separations. The functional form used to fit the reduced matrix elements is: $\mathfrak{M} (\tau) = c_0 + c_1 \tau \,$. The top-left panel shows the fit for $p = 1 \times \frac{2 \pi}{a L}$ = 0.41 GeV and $z = a $ = 0.094 fm. The top-middle panel shows the fit for $p = 2 \times \frac{2 \pi}{a L}$ = 0.82 GeV and $z = 2a $ = 0.188 fm. The top-right panel shows the fit for $p = 2 \times \frac{2 \pi}{a L}$ = 0.82 GeV and $z = 6a $ = 0.564 fm. The bottom-left panel shows the fit for $p = 4 \times \frac{2 \pi}{a L}$ = 1.64 GeV and $z = 6a $ = 0.564 fm. The bottom-middle panel shows the fit for $p = 5 \times \frac{2 \pi}{a L}$ = 2.05 GeV and $z = 4a $ = 0.376 fm. The bottom-right panel shows the fit for $p = 6 \times \frac{2 \pi}{a L}$ = 2.46 GeV and $z = a $ = 0.094 fm.}
\eefs{mockdemocn}
%%%%%%%

%%%%%%%%%%%%%%%%%%%%%
%%%%%%%%%%%%%%%%%%%%%
%%%%%%%%%%%%%%%%%%%%%
\section{Determination of gluon PDF and comparison with phenomenological distribution}\label{sec:pdf_calc}

Determining PDFs from lattice calculations involves the challenge of how best  to  extract  a  continuous distribution  from  the discrete lattice data,  compounded  by  a limited  number  of  data  points due  to  a  finite  range  of  field  separations  and  hadron momenta, and therefore a finite range of $\nu$. By performing a phenomenological analysis of the NNPDF unpolarized gluon PDF~\cite{Ball:2017nwa}, it has been found in~\cite{Sufian:2020wcv} that a $\nu$-range that is much larger than the present calculation, or any available lattice QCD determination of the gluon ITD~\cite{Fan:2020cpa,Fan:2021bcr}, is necessary to  determine the gluon distribution in the entire $x$-region from the ITD data. Therefore, we do not expect a proper determination of the gluon distribution in the entire $x$-region, especially in the small-$x$ domain. However, given our lattice data in a limited region, namely $\nu \in [0,7.07]$, we extract the gluon PDF from the reduced pseudo-ITD using the Jacobi polynomial parameterization proposed in~\cite{Karpie:2021pap}. The details of this procedure are presented in~\cite{Karpie:2021pap, Egerer:2021ymv}; here we start with the simplest form for the PDF containing the matching kernel and the leading PDF behavior, which we label as $\big[$2-param (Q)$\big]$
\bea \label{eq:2paramsQ}
\mathfrak{M} (\nu, z^2) = \int_0^1 dx \; \mathcal{K} (x \nu, \mu^2 z^2) \, \frac{x^\alpha \, (1-x)^\beta}{B(\alpha+1, \beta+1)} \, .
\eea
Here, $\mathcal{K} (x \nu, \mu^2 z^2)$ is the matching kernel that factorizes the reduced pseudo-ITD directly to the gluon PDF and the beta function, $B(a, b) = \int_0^1 r^{a-1} \, (1-r)^{b-1} \, dr$ .  To assess our fit model, and the associated systematic uncertainties, we add terms to the model. We consider the effect of adding one transformed Jacobi polynomial to the functional form of the PDF and label this model $\big[$3-param (Q)$\big]$,
\bea
    \mathfrak{M} (\nu, z^2) = \int_0^1 dx \; \mathcal{K} (x \nu, \mu^2 z^2) \; x^\alpha \, (1-x)^\beta \bigg( \frac{1}{B(\alpha+1, \beta+1)} + d_1^{(\alpha,\beta)} \, J_1^{(\alpha,\beta)} (x)  \bigg) \, .
\eea

Finally, we consider a model that we denote $\big[$2-param (Q)$\, + \, \rm P_1 \big]$ for which we add a nuisance term to capture possible $\mathcal{O} \big( {a}/{|z|} \big)$ effects. This nuisance term can be parametrized by a transformed Jacobi polynomial~\cite{Karpie:2021pap}
\bea
\mathfrak{M} (\nu, z^2) = \int_0^1 dx \, \mathcal{K} (x \nu,  \mu^2 z^2) \; \, \frac{x^\alpha \, (1-x)^\beta}{B(\alpha+1, \beta+1)}
 + \bigg( \frac{a}{|z|} \bigg) \, P_1 (\nu) \, ,
\eea
where 
\bea
P_1(\nu) = p_1^{(\alpha,\beta)} \int_0^1 dx \, \cos(\nu x)\, x^\alpha (1-x)^\beta J^{(\alpha,\beta)}_1(x) \,.
\eea

The transformed Jacobi polynomials, $J^{(\alpha,\beta)}_n(x)$ are defined as,
\be
J^{(\alpha,\beta)}_n(x) = \sum_{j=0}^n \omega_{n,j}^{(\alpha,\beta)} x^j \, ,
\ee
with
\bea
\omega_{n,j}^{(\alpha,\beta)} = \binom{n}{j}  \frac{(-1)^j}{n!}   \frac{\Gamma(\alpha+n+1)\Gamma(\alpha+\beta+n+j+1)}{ \Gamma(\alpha+\beta+n+1) \Gamma(\alpha+j+1)} \, . 
\eea
Here, $\Gamma(n)$ is the Gamma function. The orthogonality relation for these transformed Jacobi polynomials becomes
\bea
\int_{0}^{1} dx\, x^\alpha (1-x)^\beta  J_n^{(\alpha,\beta)}(x) J_m^{(\alpha,\beta)}(x)   = N_n^{(\alpha,\beta)} \delta_{n,m} \, , 
\eea
where
\bea
N_n^{(\alpha,\beta)} = \frac{1}{2n + \alpha + \beta + 1} 
\frac{\Gamma(\alpha+n+1)\Gamma(\beta+n+1)}{n!\,\Gamma(\alpha+\beta+n+1)} \, .
\eea
The transformed Jacobi polynomials form a complete basis of functions in the interval [0,1], making it possible to parameterize the PDF. 

We use Bayesian analysis to extract the PDF from the reduced pseudo-ITD. We denote the set of fit parameters, which includes the exponents $\alpha$, $\beta$, and the linear coefficients of the Jacobi series for the PDF and additional terms, by $\theta$.  Bayes' theorem gives the posterior distribution, $P[\theta | \mathfrak{M}, I]$, which describes the probability distribution of a given set of parameters being the true parameters for a given set of data, $\mathfrak{M}(\nu, z^2)$, and prior information, $I$, as
\bea
P [\theta | \mathfrak{M}, I ] = \frac{ P[\mathfrak{M} |\theta] P[\theta | I] }{P[\mathfrak{M} | I]} \, .
\eea
Here, $P[\mathfrak{M} |\theta ]$ is the probability distribution of the data for a given set of model parameters. The prior distribution, which describes the probability distribution of a set of parameters given some previously held information, is $P[\theta |  I]$ and $P[\mathfrak{M} | I]$ is the marginal likelihood or evidence that describes the probability that the data are correct given the previously held information.

In our parameterization, the PDF is dominated by the leading behavior $x^\alpha(1-x)^\beta$ and the other terms should be small corrections to this. Therefore, in the $\big[$3-param (Q)$\big]$ model, our prior for the PDF model parameter, $ d_1^{(\alpha,\beta)}$ is given by a normal distribution, with a mean and width of $d_0$ and $\sigma_d$, respectively. Similarly, in the $\big[$2-param (Q)$\, + \, \rm P_1 \big]$ model, we expect the parameter for the additional $P_1$ term to be a small correction to the dominant PDF and use a normal distribution as a prior. The mean and width of the distribution are given by $e_0$ and $\sigma_e$.

Guided by phenomenological fits of PDFs, we set $\alpha$ and $\beta$ to be positive and their prior distributions are set to be log-normal distributions,
\bea
P(x,\mu_l,\sigma,x_0) = \frac{1}{(x-x_0)\sigma \sqrt{2\pi}} e^{-{[\log(x-x_0) - \mu_l]^2}/{2\sigma^2}},
\eea
where $\mu_l$ is the mean and $\sigma^2$ the variance of the distribution of $\log(x-x_0)$, and $x_0$ is the lower bound of the log-normal distributions. The most likely parameters of the model are found by maximizing the posterior distribution.  This is performed by minimizing the negative log of the posterior distribution,
\be
L^2 = -2 \log(P[\theta | \mathfrak{M}, I]) + C,
\ee
where $C$ is the normalization of the posterior, which is independent of the model parameters.

In Fig.~\ref{model:comp:itd}, we compare the light-cone ITDs obtained from these three models.  Adding more terms to the functional form of the PDF or adding more nuisance terms does not improve the quality of the fits and the limited Ioffe-time range does not allow us to add an arbitrary number of parameters to the fit models. Fig.~\ref{model:comp:itd} demonstrates that the ITDs do not differ among the three models and the resulting PDFs remain quantitatively the same. We list the $L^2$/d.o.f. and $\chi^2$/d.o.f. of the models in Table~\ref{tab:l:chi} and find no significant change. The $\chi^2$/d.o.f. and $L^2$/d.o.f. values are also in the acceptable range and their proximity shows that the prior distributions on the PDF  parameters do not have a significant effect on the fit. Therefore, for our following discussion, we focus on the $\big[$2-param (Q)$\big]$ model. 

\begin{table}
  \setlength{\tabcolsep}{10pt}
  \renewcommand{\arraystretch}{1.5}
  \begin{tabular}{ccc}
  \toprule
    Model & $L^2/\rm{d.o.f.}$ & $\chi^2/\rm{d.o.f.}$\\
    \midrule
    2-param (Q) & 1.07 & 0.81 \\
    3-param (Q) & 1.11 & 0.82 \\
    2-param (Q)$\, + \, \rm P_1$ & 1.04 & 0.77 \\
    \bottomrule
  \end{tabular}
\caption{The $L^2/\rm{d.o.f.}$ and the $\chi^2/\rm{d.o.f.}$ of different models used to perform Jacobi polynomial parameterization of the lattice reduced pseudo-ITD to calculate the gluon PDF. \label{tab:l:chi}}
\end{table}

\befs 
\centering
\includegraphics[scale=0.65]{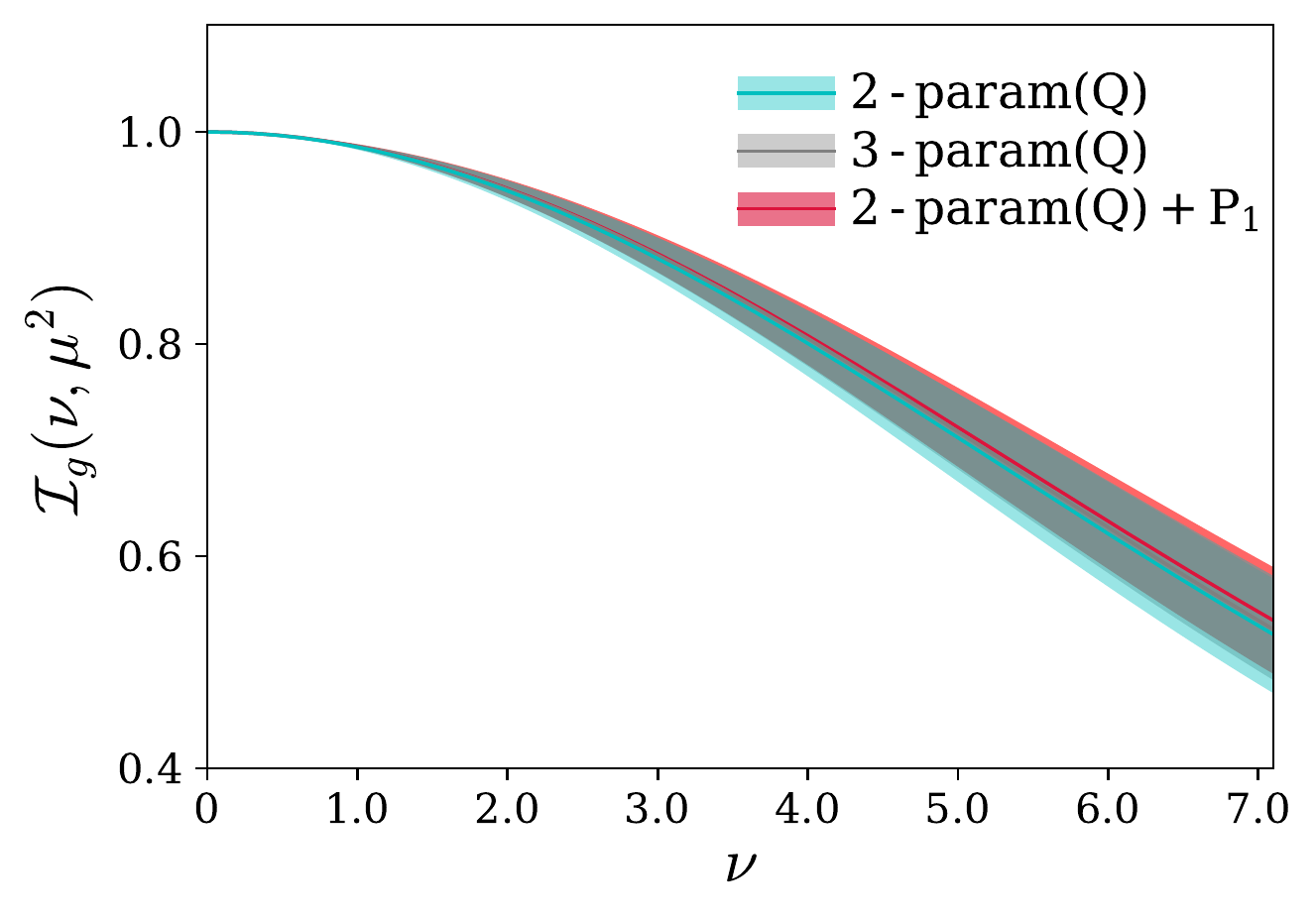}
\includegraphics[scale=0.65]{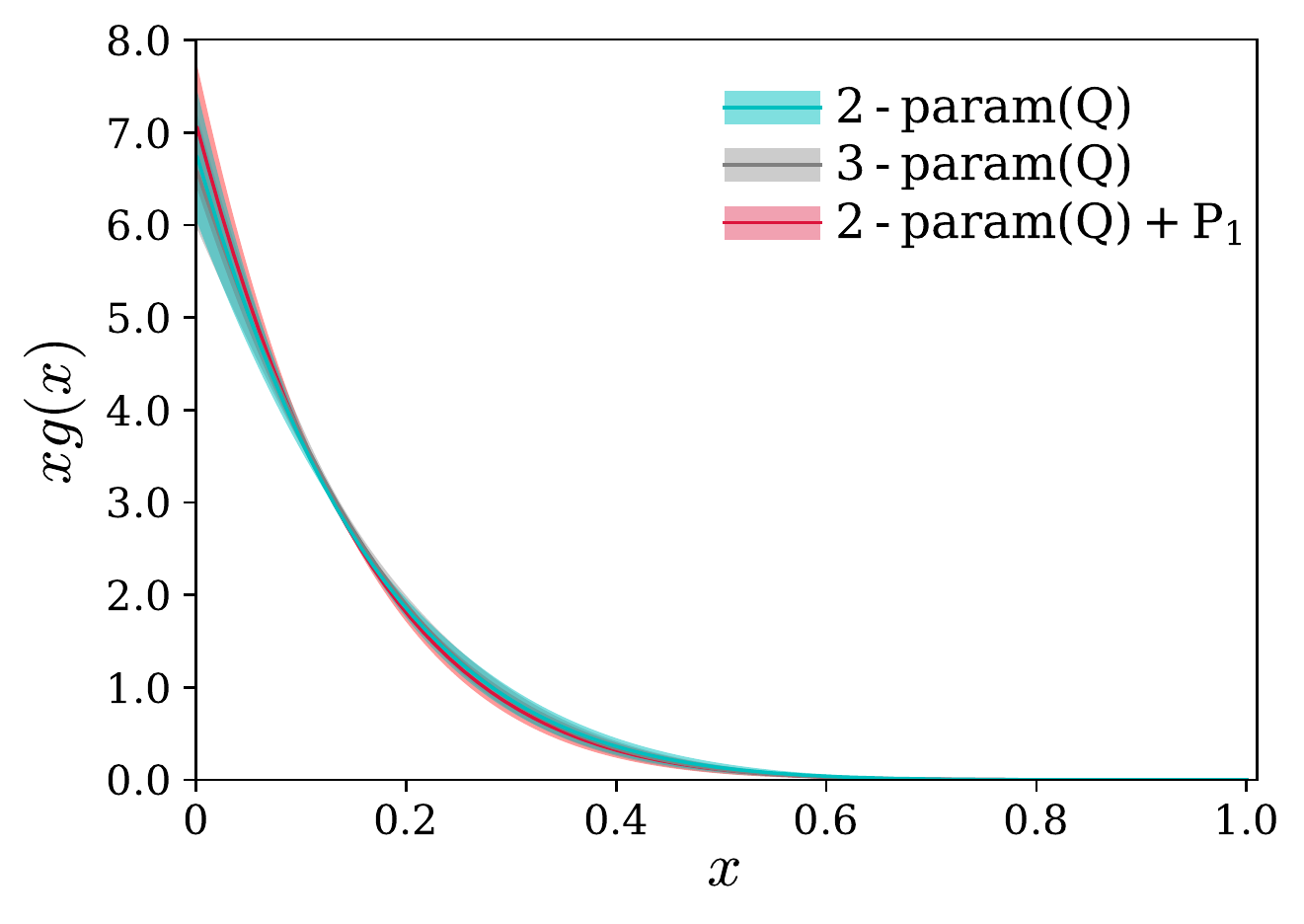}
\caption{\label{model:comp:itd} Comparison among light-cone Ioffe-time distributions calculated using Jacobi polynomial parameterization and the corresponding $x\,g(x)$ distributions at 2 GeV in the $\overline{\rm MS}$-scheme.}
\eefs{mockdemocn}

\begin{figure}[!htb]
\center{\includegraphics[scale=0.7]{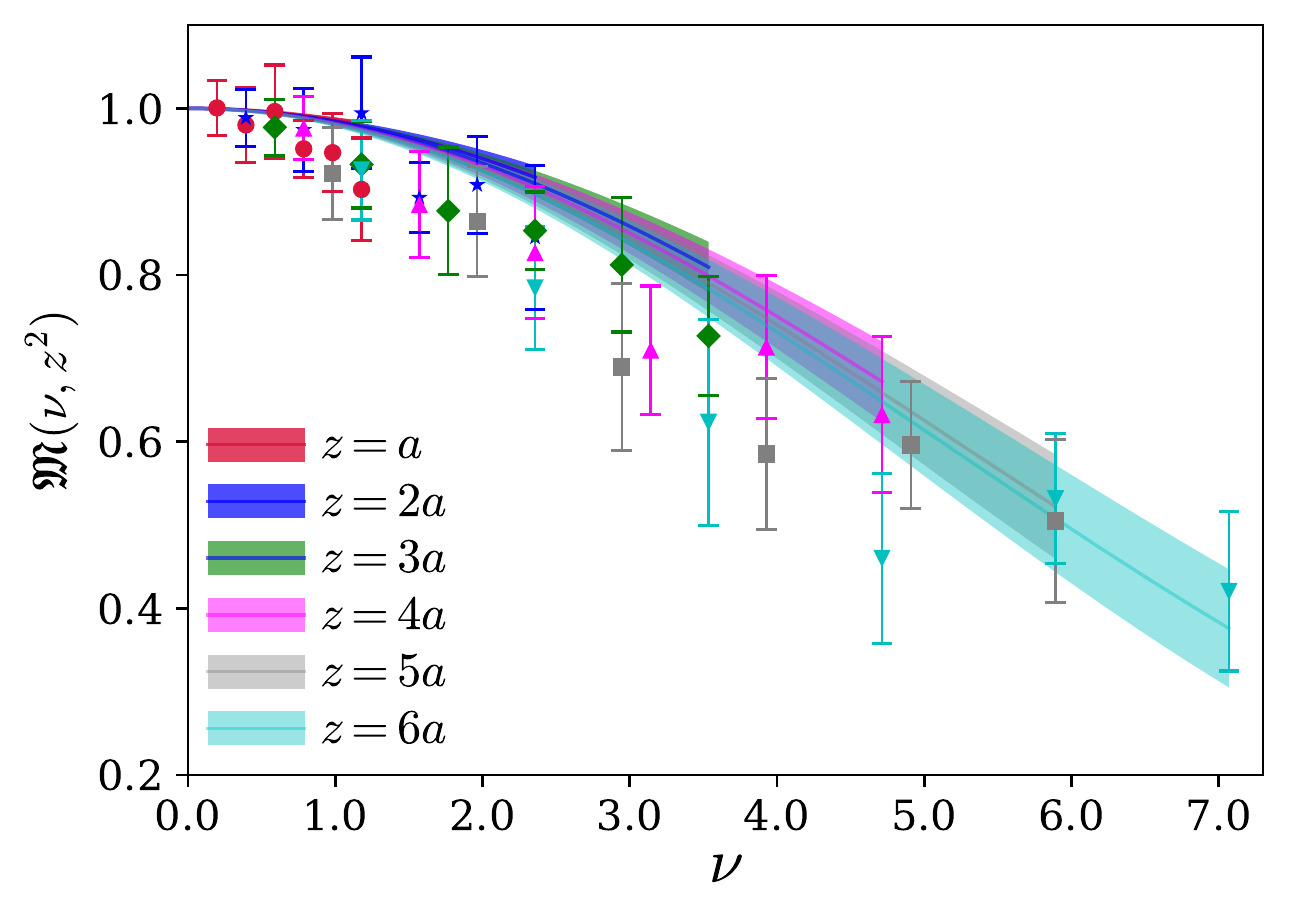}}\caption{Lattice reduced pseudo-ITD shown along with their reconstructed fitted bands calculated for the model: 2-param (Q).}\label{2param:pitd}
\end{figure}

\begin{figure}[!htb]
\center{\includegraphics[scale=0.7]{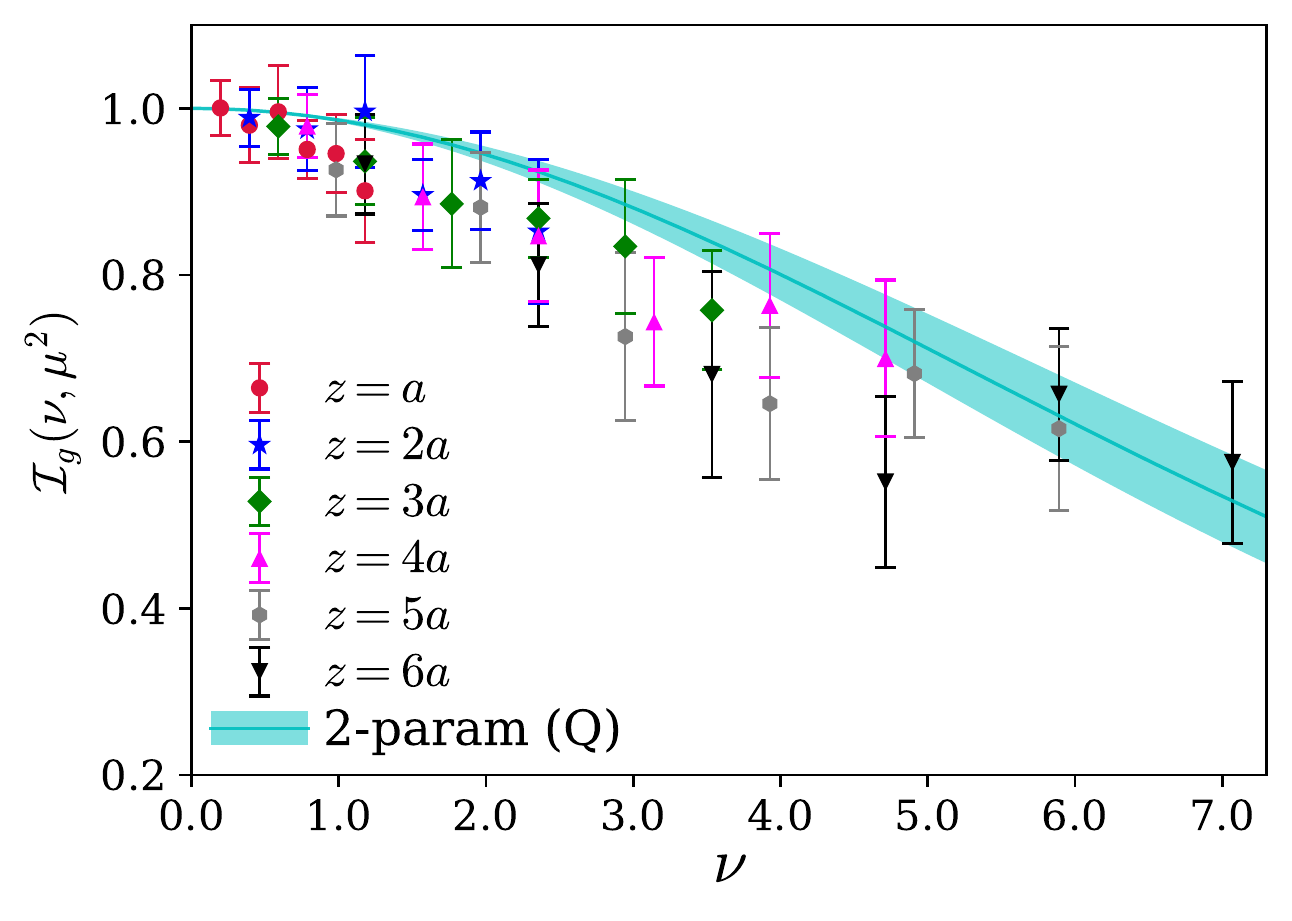}}\caption{ Ioffe-time distribution after the implementation of the perturbative matching kernel on the lattice reduced pseudo-ITD data along with the light-cone ITD calculated for the model: 2-param (Q), in the $\overline{\rm MS}$ renormalization scheme at 2 GeV.}\label{2param:itd}
\end{figure}

In Fig.~\ref{2param:pitd}, the reduced pseudo-ITD calculated is shown for different separations, $z$, along with its fitted bands obtained from the $\big[$2-param (Q)$\big]$ model. In Fig.~\ref{2param:itd}, we plot the light-cone Ioffe-time distribution with the lattice data modified by the matching kernel from the short distance factorization. SDF removes the logarithmic $z^2$ dependence of the reduced pseudo-ITD, and introduces the $\mu^2$ dependence on the light-cone Ioffe-time distribution. This effect can be observed in Fig.~\ref{2param:itd}, where after applying the matching kernel, the lattice data points with different field separations shift upward, depending on their field separations, and the data points fall on a regular light-cone Ioffe-time distribution for all $z^2$. In previous pseudo-PDF calculations such as the pion valence quark distribution determination~\cite{Joo:2019bzr}, the PDF moments extracted by implementing SDF show the logarithmic $z^2$ dependence removed for $z$ up to 1 fm. Similar results can be found in~\cite{Karpie:2018zaz}, where the moments of quark distribution in the nucleon calculated through SDF are found to be independent of a logarithmic $z^2$ effect for $z$ as large as 0.93 fm.  On the other hand, if SDF breaks down, we should see a non-polynomial $z^2$ dependence in the lattice data, especially for large $z^2$. We do not see such behavior within the current statistics. Instead, the lattice data, after modification by the matching kernel, aligns with the light-cone Ioffe-time distribution band, including the large $z^2$ data points, indicating that SDF is quite successful in extracting the Ioffe-time distribution.

In Fig.~\ref{2param:pdf}, we present the unpolarized gluon PDF (cyan band) extracted from the $\big[$2-param (Q)$\big]$ model (fit Eq.~\eqref{eq:2paramsQ}) and compare this with the gluon PDFs extracted from the phenomenological data sets CT18~\cite{Hou:2019efy}, NNPDF3.1~\cite{Ball:2017nwa}, and JAM20~\cite{Moffat:2021dji} at $\mu=2$ GeV. A similar comparison can be made with the other global fits of the gluon PDF, such as with CJ15~\cite{Accardi:2016qay}, HERAPDF2.0~\cite{Abramowicz:2015mha}, MSHT20~\cite{Bailey:2020ooq}.  To determine the normalization of the gluon PDF according to Eq.~\eqref{eq:matching}, we need to normalize the extracted PDF with the gluon momentum fraction. There has been a number of lattice calculations to extract the gluon momentum fraction~\cite{Alexandrou:2020sml, Yang:2018bft}, as well as phenomenological calculations~\cite{Ball:2017nwa, Hou:2019efy}. We take the results from~\cite{Alexandrou:2020sml}, which is $\langle x \rangle_g$=0.427(92) in the $\ms$ scheme at renormalization scale $\mu = 2$ GeV, and apply this normalization to our gluon PDF. One could similarly adopt the normalization from the $\langle x \rangle_g$ determination in~\cite{Yang:2018bft}.  We consider the uncertainties of our extracted gluon PDF and the gluon momentum fraction from~\cite{Alexandrou:2020sml} to be uncorrelated and determine the total uncertainty in the PDF. The statistical uncertainty of the gluon PDF determined from the fit Eq.~\eqref{eq:2paramsQ} and the uncertainty from the normalization using $\langle x \rangle_g$ are added in quadrature and the final uncertainty is shown as the outer band in Fig.~\ref{2param:pdf}.

%%%%%
\befs 
\centering
\includegraphics[scale=0.65]{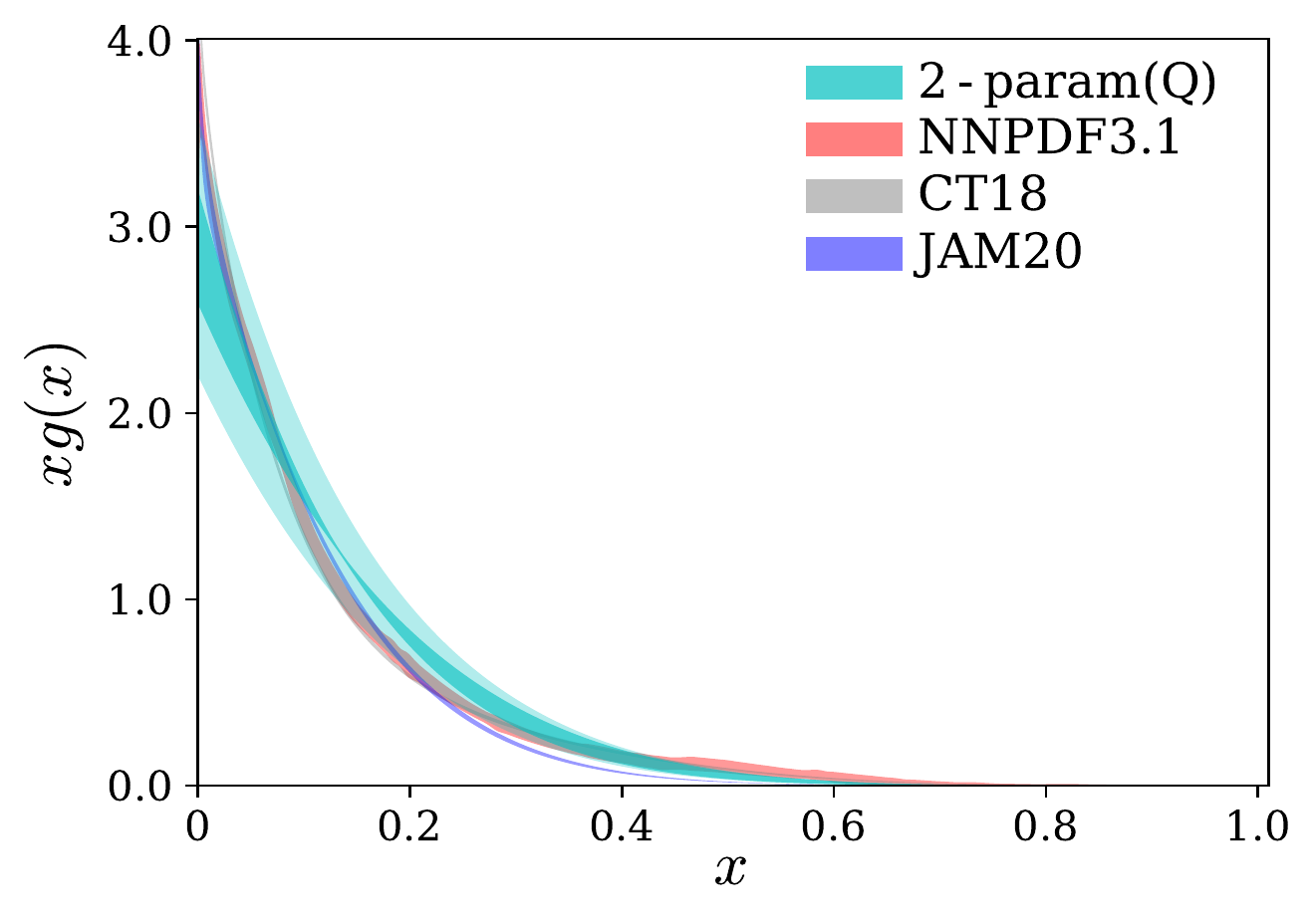}
\includegraphics[scale=0.63]{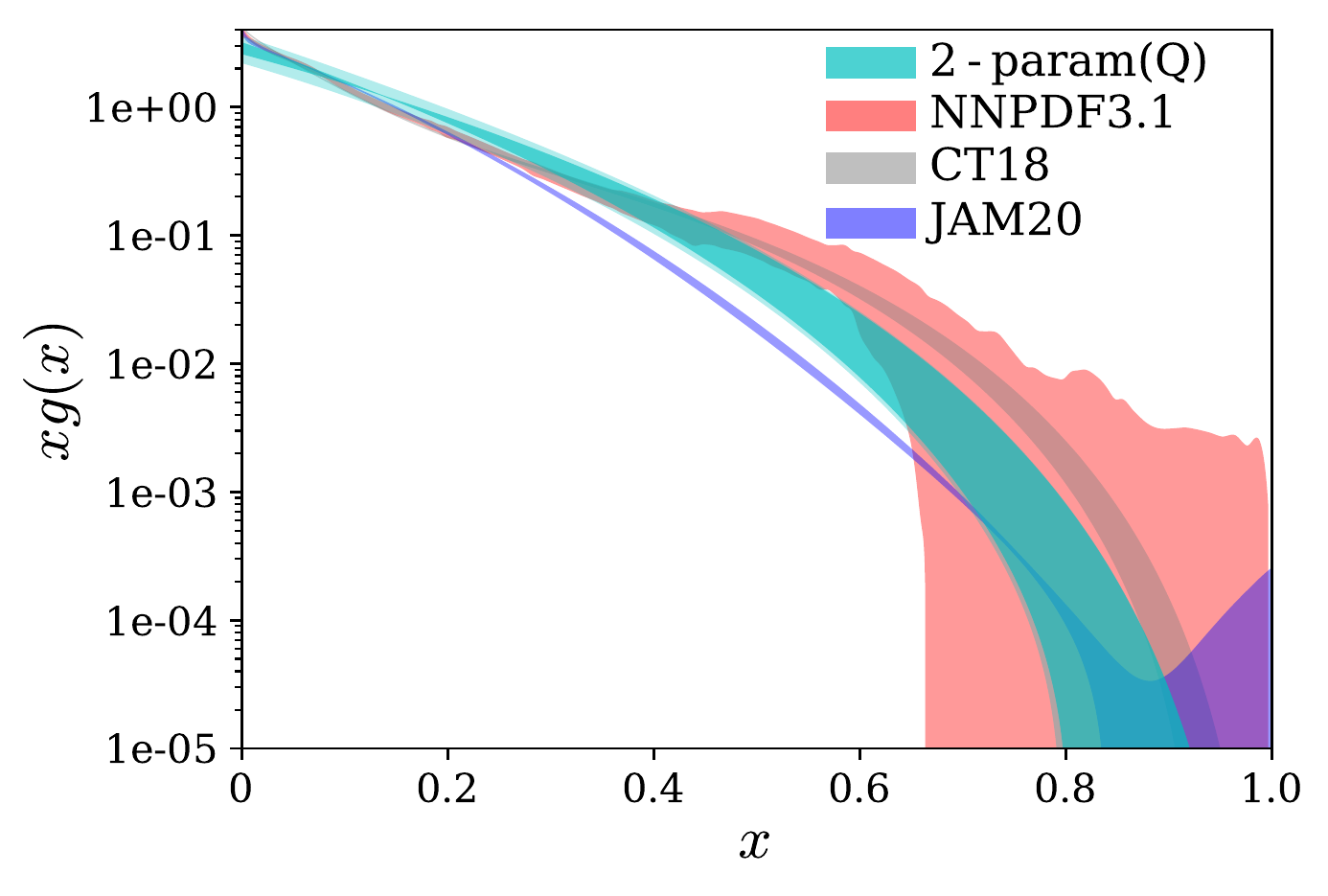}
\caption{\label{2param:pdf} Unpolarized gluon PDF (cyan band) extracted from our lattice data using the 2-param (Q) model. We compare our results to gluon PDFs extracted from global fits to experimental data, CT18~\cite{Hou:2019efy}, NNPDF3.1~\cite{Ball:2017nwa}, and JAM20~\cite{Moffat:2021dji}. The normalization of the gluon PDF is performed using the gluon momentum fraction $\langle x \rangle_g^{\ms}(\mu=2\,\mathrm{GeV})$=0.427(92) from~\cite{Alexandrou:2020sml}. The figures on left and right are the same distributions with different scales for $x\,g(x)$ to enhance the view of the large-$x$ region.}
\eefs{mockdemocn}
%%%%%%%

As discussed in~\cite{Sufian:2020wcv}, from the fitting of the ITD constructed from the NNPDF $x\,g(x)$ distribution, one needs the lattice data beyond $\nu\sim 15$ to evaluate the gluon distribution in the small-$x$ region. In the present calculation, we can extract the ITD up to $\nu\sim 7.07$. Therefore, the larger uncertainty and difference in the small-$x$ region determined from the lattice data is expected. As a cautionary remark, we also remind the readers that we have not included the mixing of the gluon operator with the quark singlet sector in the present calculation. Moreover, this calculation is performed at the unphysical pion mass and in principle, physical pion mass, continuum, and infinite volume extrapolation should be performed for a proper comparison with the phenomenological distribution. Therefore, it remains a matter of future investigation to draw a more specific conclusion about the $x\,g(x)$ distribution extracted from the lattice QCD calculation in the large-$x$ region. We also note that the shrinking of the statistical uncertainty band in the PDF near $x\sim 0.15$ results from the correlation of the PDF fit parameters. This feature has also been seen in previous works~\cite{Joo:2019bzr,Gao:2020ito,Fan:2020cpa,Bhat:2020ktg}.

However, within these limitations, we find the large-$x$ distribution is in reasonable agreement with the global fits of $x\,g(x)$ distribution, as can be seen from Fig.~\ref{2param:pdf}. The value of $\beta=5.85(72)$ determined in this calculation is statistically in good agreement with the leading $(1-x)^\beta$ behavior obtained in~\cite{Sufian:2020wcv} from the fit to the NNPDF3.1 gluon distribution and a recent phenomenological calculation~\cite{deTeramond:2021lxc}. The $\mathcal{I}_S(\nu,\mu^2)$ distribution, which we have not included in the present work, is expected to have an increasingly larger effect as $\nu$ increases and is expected to have an observable effect in the small-$x$ gluon distribution. However, in the present lattice calculation at heavier up- and down-quark masses, one expects the singlet distribution to increase at a slower rate compared to the phenomenological singlet distribution, therefore having a smaller effect on the Ioffe-time distribution in the $0\leq \nu \leq 7.07$-range.

%%%%%%%%%%%%%%%%%%%%%
%%%%%%%%%%%%%%%%%%%%%
\section{Conclusion and outlook}\label{conclusion}
%%%%%%%%%%%%%%%%%%%%%
%%%%%%%%%%%%%%%%%%%%%

In this paper, we present the unpolarized gluon parton distribution using the pseudo-PDF approach. We employ the distillation technique, combined with momentum smearing in our lattice. Distillation allows us not only to improve the sampling of the lattice but also to construct the nucleon two-point correlators with an extended basis of interpolators, which is necessary for the implementation of the sGEVP method. By using momentum smearing, momentum as high as 2.46 GeV is achieved. The sGEVP method combines the features of the summation method and GEVP technique, suppressing the excited-state contributions to the matrix elements significantly. Gradient flow reduces the UV fluctuations from the flowed matrix elements. The combination of these techniques enable us to control the signal-to-noise issues to a great extent. The reduced pseudo-ITD is calculated from the flowed reduced matrix elements by fitting the $\tau$-dependence using a linear form and extrapolating to $\tau \to 0$ limit. Using the Jacobi polynomial parameterization, the gluon parton distribution is extracted directly from the reduced pseudo-ITD. Although systematics like higher-twist contributions, lattice spacing errors, infinite volume effects, unphysical pion mass effects are not refined from the parton distribution, and quark-gluon mixing is excluded from the calculation, the resultant ITD has a well-regulated signal-to-noise ratio. The gluon PDF extracted is remarkably consistent with that extracted from the phenomenological distributions. Future endeavors include performing the calculation with a larger number of gauge configurations on the same ensemble and also perform a lattice calculation of the gluon momentum fraction, which will enables us to address the systematic uncertainties more completely along with better statistics. Incorporating the quark-gluon mixing to the calculation is another task we are aiming to undertake. When all the systematic uncertainties are properly quantified and the mixing with the isoscalar quark PDF are included, the lattice calculations will help constrain the gluon PDF at large-$x$, where the PDF is less constrained by experimental data.

%%%%%%%%%%%%%%%%%%%%%%%%%%%%%%%%%%%%%%%%%%
\section{Acknowledgement}

We would like to thank all the members of the HadStruc collaboration for fruitful and stimulating exchanges. TK and RSS acknowledge Luka Leskovec and Archana Radhakrishnan for offering their generous help, which greatly assisted this research. TK is support in part by the Center for Nuclear Femtography grants C2-2020-FEMT-006, C2019-FEMT-002-05. TK, RSS, and KO are supported  by U.S. DOE Grant \mbox{\#DE-FG02-04ER41302.} AR and WM are also supported  by U.S. DOE Grant \mbox{\#DE-FG02-97ER41028.} JK is supported
by U.S. DOE grant \mbox{\#DE-SC0011941}. This work is supported by the U.S. Department of Energy, Office of Science, Office of Nuclear Physics under contract DE-AC05-06OR23177. Computations for this work were carried out in part on facilities of the USQCD Collaboration, which are funded by the Office of Science of the U.S. Department of Energy. This work was performed in part using computing facilities at The College of William and Mary which were provided by contributions from the National Science Foundation (MRI grant PHY-1626177), and the Commonwealth of Virginia Equipment Trust Fund. This work used the Extreme Science and Engineering Discovery Environment (XSEDE), which is supported by National Science Foundation grant number ACI-1548562. Specifically, it used the Bridges system, which is supported by NSF award number ACI-1445606, at the Pittsburgh Supercomputing Center (PSC) \cite{6866038, Nystrom:2015:BUF:2792745.2792775}. In addition, this work used resources at NERSC, a DOE Office of Science User Facility supported by the Office of Science of the U.S. Department of Energy under Contract \#DE-AC02-05CH11231, as well as resources of the Oak Ridge Leadership Computing Facility at the Oak Ridge National Laboratory, which is supported by the Office of Science of the U.S. Department of Energy under Contract No. \mbox{\#DE-AC05-00OR22725}. The software codes {\tt Chroma} \cite{Edwards:2004sx}, {\tt QUDA} \cite{Clark:2009wm, Babich:2010mu} and {\tt QPhiX} \cite{QPhiX2} were used in our work. The authors acknowledge support from the U.S. Department of Energy, Office of Science, Office of Advanced Scientific Computing Research and Office of Nuclear Physics, Scientific Discovery through Advanced Computing (SciDAC) program, and of the U.S. Department of Energy Exascale Computing Project. The authors also acknowledge the Texas Advanced Computing Center (TACC) at The University of Texas at Austin for providing HPC resources, like Frontera computing system~\cite{10.1145/3311790.3396656} that has contributed to the research results reported within this paper.
We acknowledge PRACE (Partnership for Advanced Computing in Europe) for awarding us access to the high performance computing system Marconi100 at CINECA (Consorzio Interuniversitario per il Calcolo Automatico dell’Italia Nord-orientale) under the grant Pra21-5389.  JLAB-THY-21-3469.

\bibliography{GluonITD.bib}

%%%%%%%%%%%%%%%%%%%%%%%%%%
\appendix
%%%%%%%%%%%%%%%%%%%%%%%%%%

\section{Implement of sGEVP}\label{sgevp:appendix}

In sGEVP~\cite{Bulava:2011yz, Blossier:2009kd} method, the summation method~\cite{Bouchard:2016heu} and GEVP method~\cite{Edwards:2011jj} are combined together. In order to achieve that, we construct the summed three-point correlator by summing over the three-point correlators that have the same source-sink separations, but gluonic currents inserted at different time-slices between the source and sink. From the sum, to avoid contact contributions, we exclude the three-point correlators which have gluonic currents inserted at the source time-slice or sink time-slice themselves. We construct the summed three-point correlators for different interpolator combinations at the source and the sink.
\be
C^{s}_{3pt}(t_{src}, t_{snk}) = \sum_{t_g = (t_{src}+1)}^{(t_{snk}-1)} C_{2pt}(t_{src}, t_{snk}) \; O_g (t_g) \, .
\ee

Here, $t_{src}$ and $t_{snk}$ are the lattice time-slices where the source and the sink are, respectively. The label "s" stands for summed. To implement the sGEVP, consider two sets of interpolators,
\bea
    \mathcal{O}_i (t) = \mathcal{O}_i^{(A)} (t) \; , \;\; i = 1 \dots N \nn \\
    \mathcal{O}_{i+N} (t) = \mathcal{O}_i^{(B)} (t) \; , \;\; i = 1 \dots N.
\eea

Expanding the path integral to first order in $\epsilon$, the combined $2N \times 2N$ matrix of the two-point correlators from these interpolators, $C_{ij}(t,\epsilon) = \langle \mathcal{O}_i (t) \mathcal{O}_j^\dagger (0) \rangle$, can be written in the simple block structure,
\bea
C(t, \epsilon) = 
  \Bigg[ {\begin{array}{cc}
   C_{2pt} (t)                           & \epsilon \, C_{3pt}^{s} (t) \\
   \epsilon \, C_{3pt}^{s \dagger} (t) & C_{2pt} (t)                \\
  \end{array} } \Bigg] + \mathcal{O} (\epsilon^2) \, .
\eea

Here, we set $C^{(A)} = C^{(B)} = C_{2pt}$. The $2N \times 2N$ GEVP equation, 
\be
    C(t, \epsilon) \rho_n (t, t_0, \epsilon) = \lambda_n (t, t_0, \epsilon) C(t_0, \epsilon) \rho_n (t, t_0, \epsilon) \, ,
\ee

can be rewritten into its components,

\bea
    \big[ C_{2pt} (t) \pm \epsilon C_{3pt}^{s} (t) \Big] \, u^\pm_n (t, t_0, \epsilon) = \lambda^{\pm}_n (t, t_0, \epsilon) \big[ C_{2pt} (t_0) \pm \epsilon C_{3pt}^{s} (t_0) \Big] \, u^\pm_n (t, t_0, \epsilon) \, ,
\eea

where 
\be
\rho_n^\pm = \frac{1}{\sqrt{2}} \Bigg( {\begin{array}{c}
u_n^\pm \\
\pm u_n^\pm 
\end{array} }\Bigg) \, . 
\ee

Taking the small $\epsilon$ limit, we can treat the summed three-point correlators as a perturbation. By by expanding the GEVP equation in $\epsilon$, we can write the effective matrix element as~\cite{Bulava:2011yz}, 

\bea
\mathcal{M}^{\mathrm{eff}, s}_{nn} (t, t_0) = - \partial_t  \Bigg\{ \frac{\bigg|\Big(u_n, \, \big[C_{3pt}^{s}(t)\lambda_n^{-1} (t, t_0) \, - \, C_{3pt}^{s}(t_0)\big] u_n\Big)\bigg|}{\big(u_n, \, C_{2pt}(t_0) u_n\big)} \Bigg\} \, .
\eea
Here, 
\be
\Big(u_n, \, C_{2pt}(t_0) u_n \Big) \equiv u_n^\dagger \, \big( C_{2pt}(t_0) u_n\big) \, ,
\ee
and $n$ is the index of the interpolator. In the small $\epsilon$ limit, $u_n$  and $\lambda_n (t, t_0)$ are the generalized eigenvector and the principal correlator of the generalized eigenvalue problem for the two-point correlator matrix.
\be
C_{2pt} (t) \, u_{n} (t, t_0) = \lambda_n (t, t_0) \, C_{2pt} (t_0) \, u_{n} (t, t_0) \, .\label{eq:gev}
\ee

The generalized eigenvector, $u_n$, satisfies the orthogonality condition:
\be
u_{n^\prime}^{\dagger} (t, t_0) \, C_{2pt} (t_0) \, u_{n} (t, t_0) = \delta_{n, n^\prime}.
\ee

In GEVP, we rotate the two-point correlator matrix to be diagonal in the generalized eigenvector space, eliminating the excited-state contributions significantly. In sGEVP, we rotate the summed three-point correlator matrix with the same angle by which the two-point correlator matrix is rotated to be diagonal. This rotation suppresses the excited-state contributions in the summed three-point correlators too. As the orthogonality of the generalized eigenvectors are defined with respect to $t=t_0$, the ratio of the $C_{3pt} (t)$ matrix to the principal correlator matrix, $\lambda (t, t_0)$ is ill-defined at $t=t_0$. We subtract $C_{3pt} (t_0)$ from the ratio for all $t$ to avoid this issue. 

To extract the matrix element from $\mathcal{M}^{\mathrm{eff}, s}_{nn} (t, t_0)$, we recall from the degenerate  perturbation theory that the matrix element is the first derivative of the energy with respect to the perturbation taken in the $\epsilon \rightarrow 0$ limit. Now, the effective energy is given in terms of the principal correlator~\cite{Luscher:1990ck},
\be
E_n^{\mathrm{eff}} (t, t_0, \epsilon) = - \partial_t \, log(\lambda_n (t, t_0, \epsilon)).
\ee

So, the effective matrix element can be expressed as,
\bea 
\mathcal{M}^{\mathrm{eff}, s}_{nn} (t, t_0) \equiv \frac{d}{d \epsilon} E_n^{\mathrm{eff}} (t, t_0, \epsilon) \bigg|_{\epsilon = 0} = \, \mathcal{M}_{nn} + O \big(\Delta E_{N+1, n} \; t \; \mathrm{exp}(-\Delta E_{N+1, n} \, t)\big) \, .\label{sGEVP_fit}
\eea
Here, $N$ is the total number of states.

%%%%%%%%%%%%%%%%%%%%%%%%%%
%%%%%%%%%%%%%%%%%%%%%%%%%%
\section{Zero Flow Time Extrapolated Reduced Matrix Elements}\label{zero_flow_time_reduced_mtx_elem}

For each nucleon momentum and each field separation, the flowed reduced matrix elements for different flow times are fit to a linear expression: $\mathfrak{M}(\tau) = c_0 + c_1 \tau$, where the fit parameter, $c_0$ gives the reduced pseudo-ITD at zero flow time limit. The fit parameters, $c_0$ and $c_1$ are tabulated in Table~\ref{tab:zft_reduced_mtx_elem}, along with the goodness of the fits, $\chi^2/{\rm d.o.f.}$.

%%%%%%%%%%%%%%%%%%%%%%%
%\begin{longtable}{p{0.15\textwidth} p{0.15\textwidth} p{0.20\textwidth} p{0.20\textwidth} p{0.1\textwidth}}
 \begin{table*}
  \centering
  \renewcommand{\arraystretch}{1.5}
  \setlength{\tabcolsep}{10pt}

  \begin{tabular}{ccccccccc}
    \toprule
    $p$ (GeV)  & $z \, (a)$ & $\nu$ & $c_0$ & $c_1$ & $\chi^2/{\rm d.o.f.}$ \\
    \midrule
    $0.41$ & $1$ & $0.20$ & 1.0005(328) & -0.0026(106) & 0.335 \\
    $0.41$ & $2$ & $0.39$ & 0.9885(341) & -0.0057(121) & 0.505 \\
    $0.41$ & $3$ & $0.59$ & 0.9773(338) & 0.0015(142) & 0.262 \\
    $0.41$ & $4$ & $0.79$ & 0.9765(380) & -0.0004(142) & 0.271 \\
    $0.41$ & $5$ & $0.98$ & 0.9218(553) & 0.0120(232) & 0.323 \\
    $0.41$ & $6$ & $1.18$ & 0.9260(599) & 0.0099(189) & 0.401 \\
    $0.82$ & $1$ & $0.39$ & 0.9800(448) & -0.0036(155) & 0.127 \\
    $0.82$ & $2$ & $0.79$ & 0.9741(497) & 0.0006(174) & 0.436 \\
    $0.82$ & $3$ & $1.18$ & 0.9326(522) & 0.0073(217) & 0.107 \\
    $0.82$ & $4$ & $1.57$ & 0.8847(633) & 0.0292(240) & 0.306 \\
    $0.82$ & $5$ & $1.96$ & 0.8641(658) & 0.0076(269) & 0.181 \\
    $0.82$ & $6$ & $2.36$ & 0.7843(735) & 0.0171(252) & 0.658 \\
    $1.23$ & $1$ & $0.59$ & 0.9962(558) & -0.0043(209) & 0.117 \\
    $1.23$ & $2$ & $1.18$ & 0.9945(671) & -0.0160(292) & 0.119 \\
    $1.23$ & $3$ & $1.77$ & 0.8770(766) & 0.0175(299) & 0.155 \\
    $1.23$ & $4$ & $2.36$ & 0.8271(788) & 0.0202(303) & 0.131 \\
    $1.23$ & $5$ & $2.95$ & 0.6896(1004) & 0.0458(342) & 0.096 \\
    $1.23$ & $6$ & $3.53$ & 0.6232(1234) & 0.0376(322) & 0.555 \\
    $1.64$ & $1$ & $0.79$ & 0.9514(344) & -0.0014(127) & 0.569 \\
    $1.64$ & $2$ & $1.57$ & 0.8928(423) & 0.0180(148) & 0.339 \\
    $1.64$ & $3$ & $2.36$ & 0.8533(463) & 0.0127(120) & 0.209 \\
    $1.64$ & $4$ & $3.14$ & 0.7099(769) & 0.0483(249) & 0.130 \\
    $1.64$ & $5$ & $3.93$ & 0.5853(906) & 0.0581(278) & 0.319 \\
    $1.64$ & $6$ & $4.71$ & 0.4599(1015) & 0.0801(336) & 0.470 \\
    $2.05$ & $1$ & $0.98$ & 0.9468(465) & 0.0046(163) & 1.285 \\
    $2.05$ & $2$ & $1.96$ & 0.9081(585) & 0.0119(234) & 0.107 \\
    $2.05$ & $3$ & $2.95$ & 0.8121(805) & 0.0268(373) & 0.087 \\
    $2.05$ & $4$ & $3.93$ & 0.7137(860) & 0.0196(283) & 0.258 \\
    $2.05$ & $5$ & $4.91$ & 0.5958(762) & 0.0374(238) & 0.112 \\
    $2.05$ & $6$ & $5.89$ & 0.5314(780) & 0.0431(243) & 0.274 \\
    $2.46$ & $1$ & $1.18$ & 0.9027(617) & 0.0088(162) & 0.871 \\
    $2.46$ & $2$ & $2.36$ & 0.8452(866) & 0.0262(297) & 0.320 \\
    $2.46$ & $3$ & $3.53$ & 0.7268(713) & 0.0336(235) & 0.595 \\
    $2.46$ & $4$ & $4.71$ & 0.6327(935) & 0.0313(344) & 0.050 \\
    $2.46$ & $5$ & $5.89$ & 0.5048(974) & 0.0442(300) & 0.343 \\
    $2.46$ & $6$ & $7.07$ & 0.4203(954) & 0.0349(270) & 0.388 \\
    \bottomrule
  \end{tabular}

  \caption{Reduced matrix elements extrapolated to zero flow time. The flowed reduced matrix elements are fitted using a linear form: $\mathfrak{M}(\tau) = c_0 + c_1 \tau$, where $c_0$ is the reduced matrix elements at $\tau \to 0$ . \label{tab:zft_reduced_mtx_elem}}
\end{table*}
%\end{longtable}

\end{document}